\documentclass[aps,prd,10pt,nofootinbib,twocolumn,eqsecnum,showpacs,showkeys,superscriptaddress,preprintnumbers,altaffilletter]{revtex4-1}
\usepackage{graphicx}
\usepackage{amsmath}
\usepackage{multirow}
\usepackage{hyperref}
\usepackage{soul}
\usepackage{graphicx}
\usepackage{dcolumn}
\usepackage{amssymb}
\usepackage{amsfonts}
\usepackage{amsbsy}
\usepackage{color}
\usepackage{rotating}
\usepackage[english]{babel}
\usepackage{multirow}
\usepackage{float}
\usepackage{cleveref}
\usepackage{aas_macros}
\usepackage{diagbox}

\usepackage{booktabs}
\newcommand{\dtoprule}{\specialrule{1pt}{0pt}{0.4pt}%
            \specialrule{0.3pt}{0pt}{\belowrulesep}%
            }
\newcommand{\dbottomrule}{\specialrule{0.3pt}{0pt}{0.4pt}%
            \specialrule{1pt}{0pt}{\belowrulesep}%
            }

\newcolumntype{x}[1]{>{\centering\arraybackslash\hspace{0pt}}p{#1}}

\begin{document}

\title{Characteristic features of gravitational wave lensing as probe of lens mass model}

\author{P. Cremonese}
\email{paolo.cremonese@usz.edu.pl}
\affiliation{Institute of Physics, University of Szczecin, Wielkopolska 15, 70-451 Szczecin, Poland}
\author{D. F. Mota}
\email{d.f.mota@astro.uio.no}
\affiliation{Institute of Theoretical Astrophysics, University of Oslo, P.O. Box 1029 Blindern, N-0315 Oslo, Norway}
\author{V. Salzano}
\email{vincenzo.salzano@usz.edu.pl}
\affiliation{Institute of Physics, University of Szczecin, Wielkopolska 15, 70-451 Szczecin, Poland}

\date{\today}

\begin{abstract}

To recognize gravitational wave lensing events and being able to differentiate between similar lens models will be of crucial importance once one will be observing several lensing events of gravitational waves per year.
In this work, we study the lensing of gravitational waves in the context of LISA sources and wave-optics regime. While different papers before ours studied microlensing effects enhanced by simultaneous strong lensing, we focus on frequency (time) dependent phase effects produced by one lens that will be visible with only one lensed image. We will show how, in the interference regime (i.e. when interference patterns are present in the lensed image), we are able to i) distinguish a lensed waveform from an unlensed one, and ii) differentiate between different lens models. In pure wave-optics, on the other hand, the feasibility of the study depends on the SNR of the signal and/or the magnitude of the lensing effect. To achieve these goals we study the phase of the amplification factor of the different lens models and its effect on the unlensed waveform, and we exploit the signal-to-noise calculation for a qualitative analysis.

\end{abstract}

\maketitle

\section{Introduction}

As the number of Gravitational Wave (GW) events grows, studying the possible lensing of such events is becoming more and more important in the (observational) cosmology field, because it permits to infer correctly the data of the source and add useful information linked to the lensing system (e.g. \cite{Sereno:2011ty,Liao:2017ioi,Cremonese:2019tgb,Cusin:2020ezb,Ezquiaga:2021lli}). One of the latest LIGO collaboration paper \cite{Abbott:2021iab} focused exactly on finding any evidence of lensing in the registered signals. Other papers before \cite{McIsaac:2019use,Hannuksela:2019kle,Li:2019osa,Dai:2020tpj,Liu:2020par} looked for lensing signature in GWs data, as well. None of them, though, gave positive results.

So far, in fact, there is no conclusive evidence of a registered lensed event \cite{Smith:2018,Abbott:2020mjq}. It is worth noting, though, that in \cite{Diego:2021fyd} the authors claim that this is due to the low binary black-hole (BBH) coalescence rate at high redshift adopted by the Ligo-Virgo Collaboration and that, replacing the priors on the time delay distribution with the empirical Quasar-based distribution translates in a signification fraction of BBH pairs being viable candidates for multiply-lensed events. 

Nonetheless, detection of lensed events is expected to happen in the near future \cite{Oguri:2018muv,Ng:2018abc,Li:2018prc,Gao:2021sxw,Sereno:2010dr,Takahashi:2003ix}. In particular, \cite{Ng:2018abc} shows that strong lensing rate of GWs produced by elliptical galaxies is $\approx0.2$ yr$^{-1}$ for LIGO and $\approx1$ yr$^{-1}$ for aLIGO, and \cite{Li:2018prc} adds that, for the nominal Einstein Telescope sensitivity, the rate of lensing events could rise to~$\sim~80$~yr$^{-1}$. 
One of the latest predictions \cite{Yang:2021abs} states that 2nd and 2.5 generation detectors would not detect any lensed event, while 3rd generation will give a number of lensed BBH systems of $\approx350$ per year. In \cite{Xu:2021bfn}, the authors claim that $\sim0.1\%$ of observed events are expected to be strongly lensed in the 3rd generation and that ET/CE will detect $\sim50$ lensed pairs per year.
Regarding LISA mission \cite{LISA2017}, in its mass and lens range, \cite{Gao:2021sxw} initially found that about $20-40 \%$ of the massive BBHs in the mass range of $10^5-10^6$ M$_\odot$ and the redshift range of $4-10$ should show detectable wave-optics effects. 
However, the authors of \cite{Gao:2021sxw} have revisited their analysis, considering more realistic scenarios, concluding that the resulted probability of detecting a lensed event in diffraction limit will not be as high as initially mentioned, but one order of magnitude smaller, i.e. $\sim1\%$~\footnote{Private communication from Zucheng Gao.}.

Up to 4 strong lensing events with signal-to-noise ratio (SNR) $\geq8$ are expected in a 5 year time span according to \cite{Sereno:2010dr}. That is approximately the same prediction of \cite{Takahashi:2003ix}, with 1 event per year for both point mass (PM) and singular isothermal sphere (SIS) lens.

In our previous works \cite{Cremonese:2018cyg,Cremonese:2019tgb,Cremonese:2021puh}, we studied the behaviour of GW lensing in the wave-optics (WO) regime, how the arrival time difference between GWs and electromagnetic (EM) signal in these events can constrain cosmological parameters, and we studied in depth the problem of mass-sheet degeneracy (MSD) in Gravitational Lensing of GWs. 
Continuing our series of papers on this topic, here, we want to focus on new methods to recognize lensed events and study whether and in which way different lens models may imprint different characteristic features on the lensed waveforms.
In particular, in the context of a LISA detection, we will study if and how phase effects can help to recognize lensed from unlensed signals and to differentiate between lens models. 

Many papers study phase effects in lensing of GWs.
A nice summary can be found in \cite{Takahashi:2003ix}, where the authors analyze PM and SIS models, showing how the phase of the amplification factors behaves. They find that the lens mass and the source position can be determined within $\sim0.1\% [(SNR)/10^3]^{-1}$ for lens masses larger than $10^8$ M$_\odot$ and $\gtrsim 10\% [(SNR)/10^3]^{-1}$ for lens masses smaller than $10^7$ M$_\odot$, with a typical SNR$~\sim 10^2-10^3$ for LISA sources.

For LIGO sources, in \cite{Ezquiaga:2020gdt} the authors concentrate on strong lensing in the geometrical-optics (GO) regime for PM lenses. They claim that, for a loud enough source, even with only one image, it may be possible to identify it as a strongly lensed image. 
In \cite{Meena:2019ate}, they find that the time-varying phase shift of microlensing, enhanced by strong lensing, could lead to detectable differences between different images produced by strong lensing. Similar studies \cite{Diego:2019lcd,Diego:2019rzc} find that stellar mass microlenses, embedded in a macromodel potential, can introduce interference distortions in strongly lensed gravitational waves and these effects can be used to constrain the fraction of dark matter in galaxies or clusters.
More recently, \cite{Mishra:2021xzz,Cheung:2020okf} studied the effects of microlensing on GWs, taking into account also phase effects.
In \cite{Mishra:2021xzz}, PM and singular isothermal ellipsoid (SIE) lenses were considered and the authors stated that, for microlensing features to be notable in GW signal, the strong lensing magnification needs to be substantial.
In \cite{Cheung:2020okf} the authors use PM and SIS lenses and show that diffraction effects are important when we consider GWs in the LIGO frequency band lensed by objects with masses $\lesssim 100$ M$_\odot$. 
Finally, \cite{Gao:2021sxw} also considers small phase effects to recognize lensed signal, in this case for LISA sources, and with a large impact parameter%$y\sim100$
, considering only SIS models. They find that, as already stated above, $\sim1\%$ of the sources should show detectable WO effects.

In this work, contrary to the papers summarised above, we show, for the first time, that, in the context of LISA signals, and with only one lensed image in the WO regime, the time-dependent phase shift induced by the lensing is useful not only to distinguish a lensed signal from an unlensed one, but also among different lens mass models. We will stress how this is possible when the phase effects are big enough and/or when the signal is strong enough.
Moreover, we will also exploit SNR calculations: in fact, a lensed and an unlensed event, or two signals lensed by different mass models, will not have the same phase. Then, since the SNR is sensible to the phase of the signal, it will be a useful tool when comparing signals from different mass models. 

This paper is organized as follows. In Sec.~\ref{ch:lens_geom} we introduce the basics of gravitational lensing. Sec.~\ref{ch:GWlensing} is about the lensing system: we present the source of GWs and the lens models. In Sec.~\ref{ch:phase_effects} we explain how phase effects act on lensing in different optical regimes; while in Sec.~\ref{ch:models} we show how different models leave different imprints in the lensed waveform, even if they have similar mass profiles. Sec.~\ref{sch:inference} is about inference of the parameters. Finally, we draw our conclusions in Sec.~\ref{ch:conclusions}.

\section{Lens geometry} \label{ch:lens_geom}

\begin{figure}[htbp]
%\subimport{}{lens_plot_2.tex}
\includegraphics[width=0.49\textwidth]{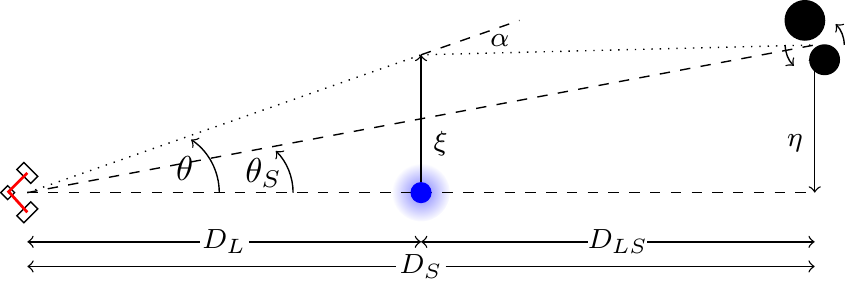}\\
\caption{Standard representation of a gravitational lensing configuration.}
\label{fig:lens_scheme}
\end{figure}

A typical lensing geometry is shown in Fig.~\ref{fig:lens_scheme}: $\vec{\theta}_{s}$ is the angular position of the source; $\vec{\theta}$ is the angular position of the image; $\vec{\hat{\alpha}}$ is the deflection angle. The angular diameter distances between observer and lens, lens and source, and observer and source are $D_{L}$, $D_{LS}$, and $D_{S}$, respectively.

As commonly in use, we will express all our relations in terms of dimensionless quantities. So, the lens equation
\begin{equation}\label{eq:lens_eq}
\vec{\beta} = \vec{\theta} - \frac{D_{LS}}{D_S}\hat{\vec{\alpha}}(\vec{\theta})\,,
\end{equation}
will translate into
\begin{equation}\label{eq:lenseq_dimless}
\vec{y} = \vec{x} - \vec{\alpha}(\vec{x}),
\end{equation}
where: $\vec{\xi}$ and $\vec{\eta}$  are the physical lengths on the lens and source plane; $\vec{x}=D_L\vec{\theta}/\vec{\xi}_0$ is the dimensionless position of the image on the lens plane; $\vec{y}=D_L\vec{\eta}/D_S \vec{\xi}_0=D_L\vec{\beta}/\vec{\xi}_0$ is the dimensionless source position on the lens plane; the scaled deflection angle is $\vec{\alpha}(\vec{x}) = \left[D_L D_{LS} / \left(\xi_0 D_S\right)\right]\hat{\vec{\alpha}}(\xi_0 \vec{x})$; and  $\vec{\xi}_0$ is a characteristic length on the lens plane.

In the following, we recall some of the main relations that we will use in our calculations. First of all, we work in the usual thin screen approximation. Therefore, instead of the volume mass density $\rho(r)$, with $r$ the three dimensional distance from the center of the lens, we work with the surface mass density, defined as \cite{Meneghetti}
\begin{equation}\label{eq:sigma}
\Sigma(\vec{\xi}) = 2 \int^{\infty}_{0} \rho(\vec{\xi},z)\,dz\, ,
\end{equation}
where $z$ is the line-of-sight direction.
We assume that the lens mass is axially distributed, thus, the two dimensional (2D) mass\footnote{In lensing theory, the 2D prefix is generally omitted because of its obviousness. Here, we specify it just to make explicit the difference with the mass that will be used to define the WO regime.} enclosed within the 2D radius $\xi$ is simply
\begin{equation}\label{eq:M2D}
M_{2D}(\xi) = 2\pi \int^{\xi}_{0} \Sigma(\xi')\xi'd\xi'\, ,
\end{equation}
and the deflection angle can be generically written as
\begin{equation}\label{eq:deflection_angle}
\hat{\alpha}(\xi) = \frac{4\,G\,M_{2D}(\xi)}{c^{2}\xi} \, .
\end{equation}
It is common to introduce another quantity, the convergence $\kappa$, defined as the dimensionless surface mass density
\begin{equation}\label{eq:convergence}
\kappa(\xi) = \frac{\Sigma(\xi)}{\Sigma_{cr}}\, ,
\end{equation}
with the critical density being
\begin{equation}\label{eq:sigma_crit}
\Sigma_{cr}= \frac{c^2}{4\pi\,G} \frac{D_S}{D_{L}D_{LS}}\, .
\end{equation}
The (2D) effective lensing potential, i.e. the 3D Newtonian potential $\Phi$ projected on the lens plane, is
\begin{equation}\label{eq:effective_pot}
\hat{\Psi}(\vec{\theta}) = \frac{D_{LS}}{D_L\,D_S} \frac{2}{c^2} \int \Phi(D_L\vec{\theta},z)dz\, ,
\end{equation}
and its dimensionless version is usually expressed as
\begin{equation}\label{eq:dimensionless_pot}
\Psi = \frac{D^{2}_{L}}{\xi^{2}_{0}} \hat{\Psi}\, .
\end{equation}
We finally remind here the relations between the scaled deflection angle and the convergence with the lensing potential, when expressed as functions of dimensionless variables \cite{Meneghetti},
\begin{eqnarray}
    \alpha(\vec{x}) &=& \vec{\nabla}_x \Psi(\vec{x})\, , \\
    \kappa(\vec{x}) &=& \frac{1}{2} \Delta_x\Psi(\vec{x})\, .
\end{eqnarray}

\section{Gravitational Wave lensing} \label{ch:GWlensing}

The lensing of GWs can be described as follows. Let $h(t)$ be the GW strain in time domain, and $\tilde{h}(f)$ the strain in the frequency domain, where $f$ is the observed GW frequency. Then, the lensed waveform will be given by \cite{Takahashi:2016jom}
\begin{equation}
    \tilde{h}_L(f) = \tilde{h}(f)\cdot F(f,\vec{y})~,
\end{equation}
where the amplification factor (AF) $F(f,\vec{y})$ is function of both the GW frequency and the dimensionless source position on the lens plane, $\vec{y}$. The AF is defined as \cite{gralen.boo}
\begin{equation}\label{eq:ampfactor}
F(w,\vec{y})=\frac{w}{2\pi i}\int \mathrm{d}^2x\exp[iwT(\vec{x},\vec{y})]~,
\end{equation}
where $w=\frac{1+z_L}{c}\frac{D_S \xi^{2}_{0}}{D_LD_{LS}} 2\pi f$ is the dimensionless GW frequency and $T(\vec{x},\vec{y})$, the dimensionless time delay
\begin{equation}\label{eq:tGOdimless}
T(\vec{x}, \vec{y})=\left[ \frac{1}{2} \left( \vec{x}-\vec{y} \right)^2 - \Psi(\vec{x}) 
\right]\, .
\end{equation}
Assuming spherical symmetry for the lens, Eq.~(\ref{eq:ampfactor}) can be written as \cite{Nakamura7}
\begin{eqnarray} \label{eq:AF}
F(w,y) &=& -iw e^{iwy^2/2} \\
&\times& \int_0^\infty \mathrm{d}x\,x\,J_0(wxy)\exp\left\{iw\left[ \frac{1}{2}x^2-\Psi(x)
\right]\right\}\,, \nonumber
\end{eqnarray}
where $J_0$ is the Bessel function of zeroth order and $y=|\vec{y}|$.
When considering phase effects, it is important to include the ``normalization" function $\phi_m$. In fact, its omission would be equivalent to assuming that we know the arrival time of the lensed signal w.r.t. the unlensed one. But, since in WO we have only one signal, we do not have such an information. 
In general, the function $\phi_m(y)$ is defined such that the time delay of the (first) image is equal to 0 \cite{Takahashi:2003ix,HerreraMartin2018}. It thus depends on the source position $y$ and on the lens mass model and can be derived from Eq.~(\ref{eq:tGOdimless}) as
\begin{equation}
    \phi_m(y) = -T(x_m,y) = -\left[\frac{1}{2}(x_m-y)^2-\psi(x_m)\right]
\end{equation}
where $x_m$ is the solution of the lens equation $y=x-\nabla_x\psi(x)$. For SIS, $\phi_m(y)=y+1/2$; while for other models the solution is generally obtained numerically. Eventually, the AF, with the normalization factor included, is
\begin{align} \label{eq:AF_norm}
F(w,y) = &-iw e^{iwy^2/2} \int_0^\infty x\,J_0(wxy) \\
&\times \exp\left\{iw\left[ \frac{1}{2}x^2-\Psi(x) + \phi_m(y)
\right]\right\}~\mathrm{d}x\,. \nonumber
\end{align}

\subsection{Gravitational Waves Source}
The source of GWs that we consider in our analysis is a  super massive binary black-hole (SMBBH) merger. The system has a (rest) total mass of $M_{tot} = 10^8~M_\odot$, $q=1$ (i.e. identical masses for the BHs, $m_1=m_2$) and redshift $z_S=1$, if not differently specified. This kind of signal would be observed by LISA \cite{LISA2017}. 

The waveforms, both in time and frequency domain, are computed using the \texttt{IMRPhenomPv3} \cite{Khan:2018fmp} approximant of the \texttt{PyCBC} \cite{alex_nitz_2020_4134752} software in \textsc{python}. To switch between time and frequency domain, we compute the Fourier transform using the software \texttt{FFTW}\footnote{http://www.fftw.org/}.

The GW source system parameters have been chosen both because we want to study lensing events in the WO and interference regimes, and because we want to abandon the microlensing regime and, instead, consider extended systems as lenses, thus using mass models different from a simple PM. Therefore, we first choose a realistic (3D) mass scale $\sim 10^{9}$ M$_{\odot}$, which can be described by a mass model more complicated than the PM. 
Then, we can set the frequency range of the event, and, eventually, the total rest mass of the source, by using the WO regime condition given by \cite{Takahashi:2016jom}
\begin{equation} \label{eq:GO_cond2}
    M_{3D,L}\leq10^5 M_{\odot}\left[\frac{(1+z_L)f}{\text{Hz}}\right]^{-1},
\end{equation}
where $M_{3D,L}$ and $z_L$ are the 3D mass and redshift of the lens, respectively.

\subsubsection{Signal-to-Noise ratio}\label{sec:SNR}
We will use the SNR as an indication of the brightness of the signal and as a method to compare two waveforms.
In a match filtering analysis, such as the one used for the detection of GWs, the SNR is calculated comparing a signal $s(t)=h(t)+n(t)$, where $h$ is the GW signal and $n$ the noise, with a template $h_{T}(t)$ \cite{maggiore2008gravitational,Ezquiaga:2020gdt}
\begin{equation} \label{eq:SNR}
\rho = \frac{\left(s|h_{T}\right)}{\sqrt{(h_{T}|h_{T})}} \approx \frac{\left(h|h_{T}\right)}{\sqrt{(h_{T}|h_{T})}}\, ,
\end{equation}
where the right-hand side is given if we neglect the correlation of the noise and the template. The inner product $(a|b)$ in the Fourier space is defined as
\begin{equation} \label{eq:inner_product}
    (a|b) = 4\,{\rm Re}\left[\int_0^\infty\frac{\tilde{a}(f)\cdot\tilde{b}^*(f)}{S_n(f)}df\right]\, ,
\end{equation}
where $\tilde{a}(f)$ is the waveform (signal) in frequency domain and $S_n(f)$ is the single-sided power spectral density \cite{LIGOScientific:2019hgc,maggiore2008gravitational}.
The optimal SNR is given when the signal matches the template, $h(t)\propto h_{T}(t)$, i.e. $\rho_{\rm opt}=\sqrt{(h|h)}$.
For our goals, we are interested in the quantity $\rho/\rho_{opt}$, i.e. the ratio between the SNR obtained by a given template w.r.t. the optimal SNR. In our calculations, we use LISA sensitivities from \cite{Cornish:2018dyw}.

To understand how well a template matches a real signal, it is necessary to calculate uncertainties. We follow a similar approach to that of \cite{Wang:2021kzt}.
In particular, the likelihood of a given GW event can be determined assuming that after the subtraction of the waveform from the signal, the noise is Gaussian \cite{maggiore2008gravitational}, i.e.
\begin{align}
\mathcal{L} &\propto \exp \left[-\frac{1}{2}(s|s)+(h|s)-\frac{1}{2}(h|h)\right] \nonumber \\
&\propto \exp \left[(h|s)-\frac{1}{2}(h|h)\right]\, ,
\end{align}
from which the $\chi^2$ is derived to be
\begin{equation}
\chi^2 = (h|h)-2(h|s) \approx \rho^{2}_{opt} \left[1-\frac{2\rho}{\rho_{opt}}\right]\,,
\end{equation}
where we do not include the common, constant term $(s|s)$, and neglect correlations between the noise and template.
Searching for a given confidence level w.r.t. to the best model corresponding to $\rho_{opt}$, translates into
\begin{equation} \label{eq:threshold}
\Delta \chi^2 \approx 2 \rho^{2}_{opt} \left[1-\frac{\rho}{\rho_{opt}}\right] \,.
\end{equation}
Given a threshold $\Delta \chi^2$, this relation allows us to determine the level of mismatch from the optimal SNR, $\rho/\rho_{opt}$, that corresponds to it. When two or three free parameters are involved in the analysis\footnote{As we shall see in Sec.~\ref{ch:lenses}, for a SIS model the free parameters are the impact parameter, identified with $y$, and the stellar dispersion velocity $\sigma$ (which defines the mass of the lens); while the NFW models are defined by $y$, and the characteristic density and radius, $\rho_s$ and $r_s$, that will be described in the text.}, the $3\sigma$ confidence level roughly corresponds to $\Delta \chi^2\sim 11.8$ and $\Delta \chi^2\sim 14.2$ \cite{Num_Rec}, respectively.

\subsection{Lens systems}\label{ch:lenses}

Regarding the lens, we have taken into consideration many mass models, but we restrict our results to only four scenarios \cite{Navarro:1995iw,Nakamura1999, Takahashi:2003ix, Meneghetti}. We compute numerically the AF using \textsc{Mathematica}, since there is no analytical solution for the integral in Eq.~\eqref{eq:AF_norm} for any of them.

The most difficult task is to choose the models in a consistent way. Indeed, we know that each of them responds to lensing in its own way. Thus, any peculiar feature in the GW lensing would be hardly comparable and interpreted as characteristic of a given mass model if we do not make sure to leverage any other influence and to compare ``similar'' mass estimations within a given distance. Moreover, we do not have real data to be used to constrain such models in order to describe the same observational picture.

\begin{figure*}[htbp]
    \centering
    \includegraphics[width=0.45\textwidth]{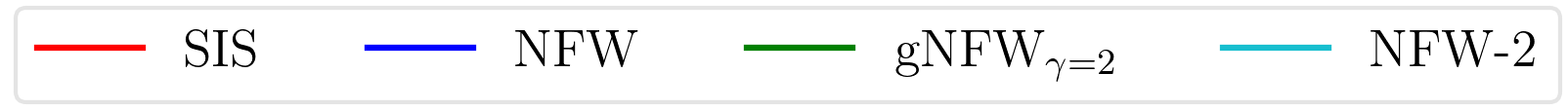}\\
    \includegraphics[width=0.48\textwidth, trim={0 0 0 1.2cm}, clip]{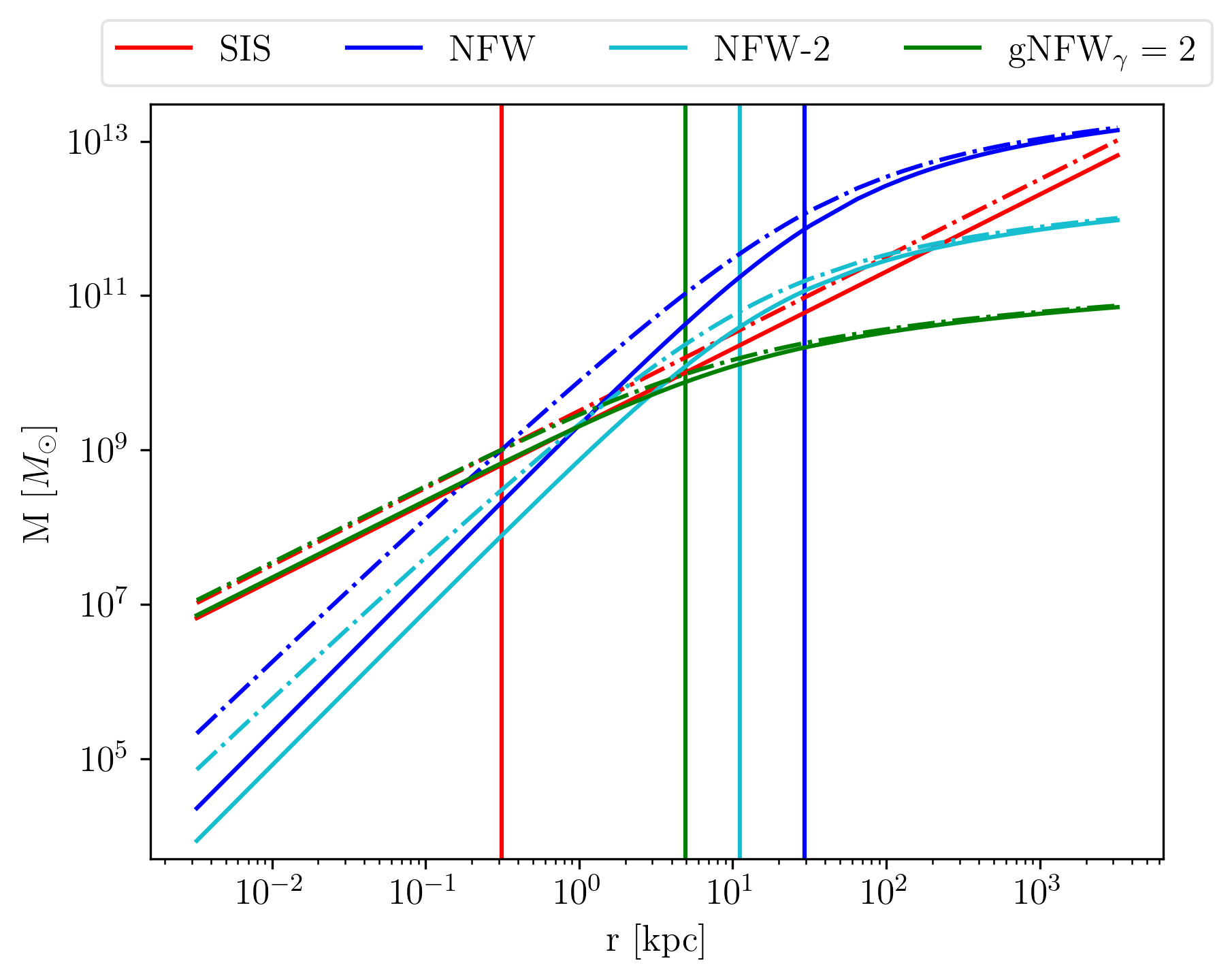}
    \includegraphics[width=0.48\textwidth, trim={0 0 0 1.2cm}, clip]{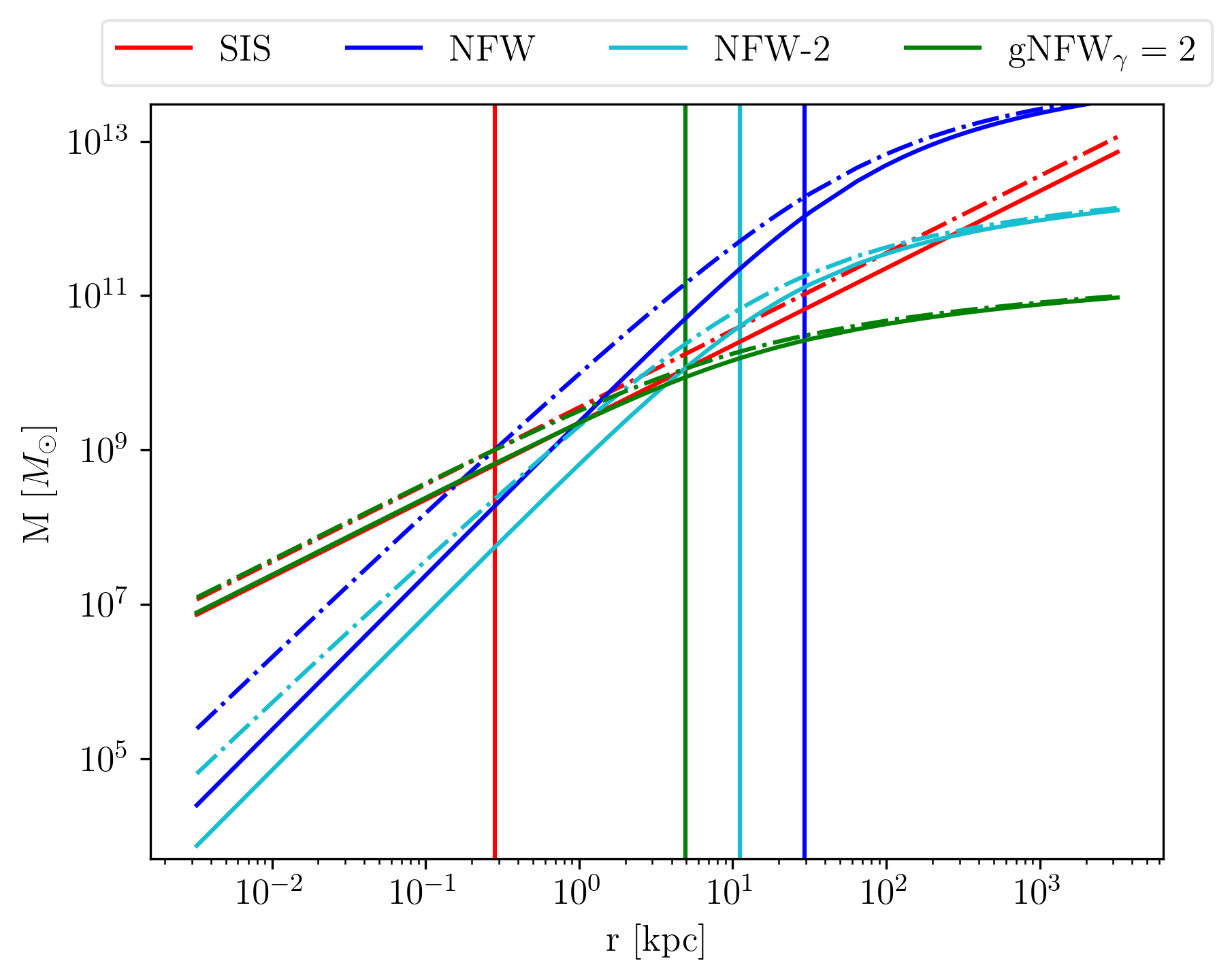} % left bottom right top 
    \caption{Mass profiles of the lens mass models that will be studied in the paper. In the \textit{left panel}, the lenses are at $z_L=0.5$, in the \textit{right panel}, at $z_L=0.15$. Solid lines are 3D masses and dot-dashed lines are 2D ones. The parameters used for each model are shown in Table~\ref{tab:results9}. The vertical lines show the physical characteristic length in the lens plane, $\xi_0$, corresponding to the impact parameter $y=1$, for each model.}
    \label{fig:masses}
\end{figure*}

The only constrain we have is from Eq.~\eqref{eq:GO_cond2}, which tells us that we can achieve WO when the 3D lens mass satisfied the condition $M_{3D,L} \leq 10^9$ $M_\odot$. This single condition, of course, brings a lot of degeneracy when it comes to fixing the various mass models. The details of our choices are explained below and summarized in Table~\ref{tab:results9}. Eventually, we get mass profiles that are very similar to each other, as Fig.~\ref{fig:masses} shows. The characteristic strains in the frequency domain for the cases taken into consideration in this study are shown in Fig.~\ref{fig:FDs}. Compatibly with LISA sensitivity, if not differently specified, the GW source is located at redshift $z_S=1$, and we choose two cases for the lens system, one at $z_L=0.5$ and the other at $z_L=0.15$.
{\renewcommand{\tabcolsep}{1.5mm}
{\renewcommand{\arraystretch}{1.5}
\begin{table*}[htbp]
\centering
\caption{Parameters of the lens models considered in this work. Rows are divided into four groups: group $1$: name of the model; group $2$: we explicitly specify the input parameters and conditions from which all other quantities are derived; group $3$: we provide the main characteristic parameters of each model; group $4$: additional parameters that might be helpful to frame the mass behaviours. In all cases we consider a cosmological background with $H_0=74$ km s$^{-1}$ Mpc$^{-1}$, $\Omega_m=0.3061$, $z_L=0.15$ and $z_S=1$.}
\resizebox*{\textwidth}{!}{
\begin{tabular}{c||c||c||c||c}
\dtoprule

 \multicolumn{5}{c}{$z_L=0.5$} \\
 \hline
  & SIS & NFW & NFW-2 & gNFW $\gamma=2$ \\
 \hline
 & $M_{2D}(\theta^{SIS}_E D_L)=10^9$ $M_\odot$ & $M_{2D}(\theta^{SIS}_E D_L)=10^9$ $M_\odot$ & $M_{2D}(r_{200})=M_{2D,SIS}(r_{200})$ &  $M_{2D}(\theta^{SIS}_E D_L)=10^9$ \\
  &                              & $c_{200}=10$ & $c_{200}=10$ & $c_{200}=10$ \\
\hline 
$\sigma$ (km/s)      & $66.26$ & $-$ & $-$ & $-$ \\
$\rho_s$ (kg/m$^3$)  & $-$     & $7.95\cdot10^{-22}$ & $7.95\cdot10^{-22}$ & $4.94\cdot10^{-22}$ \\
$r_s$ (kpc)          & $-$  & $29.38$ & $11.12$ & $4.92$ \\
\hline
$\theta_E$ (arcsec)  &  $2.62\cdot10^{-7}$ &  $2.62\cdot10^{-7}$ & $-$ & $2.62\cdot10^{-7}$ \\ 
$\theta_E D_L$ (kpc) & $0.312$ &  $0.312$ & $-$ & $0.312$  \\
$M_{2D}(\theta_E D_L)$ ($M_\odot$)  & $10^9$ & $10^9$ & $-$ & $10^{9}$ \\
$M_{3D}(\theta_E D_L)$ ($M_\odot$)  & $6.37\cdot 10^8$ & $2.08\cdot 10^8$ & $-$ & $6.71\cdot 10^8$\\
$M_{2D}(r_s)$ ($M_\odot$)       & $-$ & $1.15\cdot10^{12}$ & $6.22\cdot10^{10}$ & $9.58\cdot10^{9}$ \\
$M_{3D}(r_s)$ ($M_\odot$)       & $-$ & $7.23\cdot10^{11}$ & $3.92\cdot10^{10}$ & $7.56\cdot10^{9}$\\
%$c_{200}$                  & $-$ & $10$ & $10$ & $10$ \\
$r_{200}$ (kpc)            & $-$ & $293.8$ & $111.16$ & $49.2$ \\
$M_{2D,200}$ ($M_\odot$)      & $-$ & $6.58\cdot10^{12}$ & $3.56\cdot10^{11}$ & $2.93\cdot10^{10}$ \\
$M_{3D,200}$ ($M_\odot$)      & $-$ & $5.58\cdot10^{12}$ & $3.02\cdot10^{11}$ & $2.62\cdot10^{10}$ \\
\hline
\hline

 \multicolumn{5}{c}{$z_L=0.15$} \\
 \hline
  & SIS & NFW & NFW-2 & gNFW $\gamma=2$ \\
 \hline
 & $M_{2D}(\theta^{SIS}_E D_L)=10^9$ $M_\odot$ & $M_{2D}(\theta^{SIS}_E D_L)=10^9$ $M_\odot$ & $M_{2D}(r_{200})=M_{2D,SIS}(r_{200})$ &  $M_{2D}(\theta^{SIS}_E D_L)=10^9$ \\
  &                              & $c_{200}=10$ & $c_{200}=10$ & $c_{200}=10$ \\
\hline 
$\sigma$ (km/s)      & $69.75$ & $-$ & $-$ & $-$ \\
$\rho_s$ (kg/m$^3$)  & $-$     & $5.34\cdot10^{-22}$ & $5.34\cdot10^{-22}$ & $3.32\cdot10^{-22}$ \\
$r_s$ (kpc)          & $-$  & $47.80$ & $14.28$ & $6.26$ \\
\hline
$\theta_E$ (arcsec)  &  $5.52\cdot10^{-7}$ &  $5.52\cdot10^{-7}$ & $-$ & $5.52\cdot10^{-7}$ \\ 
$\theta_E D_L$ (kpc) & $0.281$ &  $0.281$ & $-$ & $0.281$  \\
$M_{2D}(\theta_E D_L)$ ($M_\odot$)  & $10^9$ & $10^9$ & $-$ & $10^{9}$ \\
$M_{3D}(\theta_E D_L)$ ($M_\odot$)  & $6.37\cdot 10^8$ & $1.86\cdot 10^8$ & $-$ & $6.63\cdot 10^8$\\
$M_{2D}(r_s)$ ($M_\odot$)       & $-$ & $3.32\cdot10^{12}$ & $8.86\cdot10^{10}$ & $1.32\cdot10^{10}$ \\
$M_{3D}(r_s)$ ($M_\odot$)       & $-$ & $2.09\cdot10^{12}$ & $5.58\cdot10^{10}$ & $1.05\cdot10^{10}$\\
%$c_{200}$                  & $-$ & $10$ & $10$ & $10$ \\
$r_{200}$ (kpc)            & $-$ & $478.0$ & $142.8$ & $62.6$ \\
$M_{2D,200}$ ($M_\odot$)      & $-$ & $1.90\cdot10^{13}$ & $5.07\cdot10^{11}$ & $4.05\cdot10^{10}$ \\
$M_{3D,200}$ ($M_\odot$)      & $-$ & $1.61\cdot10^{12}$ & $4.29\cdot10^{11}$ & $3.62\cdot10^{10}$ \\
\dbottomrule
    \end{tabular}}
    %\caption{}
    \label{tab:results9}
\end{table*}
}}
\subsubsection{Singular Isothermal Sphere (SIS)}

This model is characterized by a volume density
\begin{equation}\label{eq:SIS_density}
        \rho_{SIS}(r) = \frac{\sigma^2}{2\pi\,G\,r^{2}}\, ,
\end{equation}
where $\sigma$ is the line-of-sight velocity dispersion of the components of the considered structure (stars in a galaxy, galaxies in a cluster, etc$\ldots$). The typical characteristic length $\xi_0$ on the lens plane for SIS is generally the Einstein radius, $D_L\,\theta_E$. Consequently, the (dimensionful) impact parameter is given by
\begin{equation}\label{eq:imp_par_SIS}
    y_{SIS}\cdot\xi_0 = y_{SIS}\cdot(\theta^{SIS}_{E} D_L)~.
\end{equation}
Although the SIS model is not physically sounded for the divergence of the surface density at $\xi \rightarrow 0$ and of the mass at large $\xi$, it fits quite well many observations.

Finally, we decide to closely follow what is done in \cite{Takahashi:2016jom}, even though there is no standard procedure in literature: we consider a lens with a mass within the Einstein radius of $M_{2D}(D_L\theta_E) = 10^9~M_\odot$. Note that fixing the 2D masses to such a value will automatically make the 3D masses fulfill the condition $M_{3D,L}\leq 10^{9}$ $M_{\odot}$ derived from Eq.~(\ref{eq:GO_cond2}), as the 3D masses are generally smaller than the 2D ones.
        
\subsubsection{Navarro-Frenk-White (NFW)}

This model has a density profile
        \begin{equation}\label{eq:NFW_density}
        \rho_{NFW}(r) = \frac{\rho_s}{\frac{r}{r_s}\left(1+\frac{r}{r_s}\right)^2}\, ,
        \end{equation}
where $\rho_s$ and $r_s$ are the characteristic NFW density and radius. In order to fix and define the NFW model we will work with, we prefer to work with the $\Delta$-parameters. The characteristic NFW density $\rho_s$, for example, can be expressed as
\begin{equation}\label{eq:rhos200}
            \rho_s = \frac{\Delta}{3}\rho_c\, \frac{c_{\Delta}^3}{\ln(1+c_{\Delta})-\frac{c_{\Delta}}{1+c_{\Delta}}},
\end{equation}
where the $\Delta$ means that all quantities are calculated at the radius $r_{\Delta}$, where the density of the system is $\Delta$ times the critical density of the Universe at the same redshift of the lens. In our case, we consider $\Delta=200$, the so-called virial value. The concentration parameter $c_{\Delta}$ that appears in Eq.~(\ref{eq:rhos200}) is defined as
\begin{equation}\label{eq:c200}
c_{\Delta} = \frac{r_{\Delta}}{r_s},        
\end{equation}        
where $r_{\Delta}=r_{200}$ in our case. The 3D mass within a sphere of radius $r_{\Delta}$ is finally written as \cite{Meneghetti, Salzano:2017qac}
\begin{equation}\label{eq:M200}
            M_{3D,\Delta} = \frac{4}{3} \pi r_{\Delta}^3 \Delta\rho_c 
            = 4 \pi r_{s}^3 \rho_s \left[ \ln(1+c_{\Delta}) -\frac{c_{\Delta}}{1+c_{\Delta}}\right].
\end{equation}
These parameters are to be preferred because it is well known that a clear correlation exists between $c_{200}$ and $M_{200}$ \cite{Wang:2019ftp, Gilman:2019bdm, Diemer:2018vmz}, and we will use it to build up realistic models. In particular, using Eq.~(\ref{eq:M2D}), we decided to set $M_{2D}(D_L\theta_E^{SIS})=10^9$ $M_{\odot}$: the projected mass of the lens derived from the NFW model, within a physical distance from the center equal to the SIS Einstein radius, will be $10^9~M_\odot$, which is the same value taken for the SIS model. In order to do that, we must necessarily fix at least another parameter, which we chose to be $c_{200}=10$, as this value is compatible with such masses from observations \cite{Gilman:2019bdm,Diemer:2018vmz}. These choices ensure that the SIS and NFW mass profiles, within a large range of distances, are of the same order of magnitude and not very dissimilar on global scales, as one can see in Fig.~\ref{fig:masses}. 

When working with NFW models, the characteristic scale $\xi_0$ on the lens plane is generally taken to be characteristic radius $r_{s}$. Thus, the dimensionful impact parameter is
\begin{equation}\label{eq:imp_par_NFW}
     y_{NFW}\cdot\xi_0 = y_{NFW}\cdot r_s ~ .
\end{equation}
Considering that the GW waveforms and the corresponding AF are functions of the dimensionless parameter $y$, the differences in the $y$'s among the various models must be taken into account in order to compare similar configurations, both in mass and in distance. This can be seen from Fig.~\ref{fig:masses}, where the vertical lines show the position of the impact parameter at $y=1$ for each model. Thus, for example, a SIS configuration with $y_{SIS}\sim1$ will correspond to a NFW one with $y_{NFW}\sim0.01$ for $z_L=0.5$ and $y_{NFW}\sim0.005$ for $z_L=0.15$.
  
\subsubsection{Scaled NFW (NFW-2)} 

In this case, we still have a NFW profile for the mass density, but the parameters are derived fixing differently the mass. In particular, we fix the NFW lens mass at $r_{200}$ to the same value that the SIS, previously defined, would have at the same distance from the center of the lens, i.e. $M_{2D}(r_{200})=M_{2D,SIS}(r_{200})$. The impact parameter, here, is defined as in Eq.~\eqref{eq:imp_par_NFW}, but one must take into account that the $r_s$ of this model is different from the previous NFW. In particular, we have that $y_{SIS}\sim1$ will correspond to $y_{NFW-2}\sim0.03$ for the case of lens at $z_L=0.5$, while $y_{NFW-2}\sim0.02$ for $z_L=0.15$.
    
\subsubsection{Generalized NFW} 

The mass density of the generalized NFW model is given by \cite{Keeton:2001ss}
\begin{equation}\label{eq:GNFW_density}
        \rho_{gNFW}(r) = \frac{\rho_s}{\left(\frac{r}{r_s}\right)^{\gamma}\left(1+\frac{r}{r_s}\right)^{3-\gamma}}\, .
\end{equation}
It is known that a general analytical solution for such models is not possible. With $\gamma = 1$ we recover the standard NFW model. Here, we focus on the case with $\gamma=2$:
\begin{equation}\label{eq:GNFW2_density}
        \rho_{gNFW}(r) = \frac{\rho_s}{\left(\frac{r}{r_s}\right)^2\left(1+\frac{r}{r_s}\right)}\, .
\end{equation}
For the gNFW model, once again, we use the $c_{200}-M_{200}$ relation as for the NFW case and we set the parameters so that the gNFW mass within the SIS Einstein radius is equal to the mass in the SIS model, i.e. $M_{2D}(\theta_E^{SIS} D_L)=10^9$ $M_{\odot}$. The impact parameter $y$, in this case, will be such that $y_{SIS}\sim1$ will correspond to $y_{gNFW}\sim0.06$ for a lens at $z_L=0.5$, and $y_{gNFW}\sim0.04$ for a lens at $z_L=0.15$.

\begin{figure*}[htbp]
    \centering
    \includegraphics[width=0.48\textwidth]{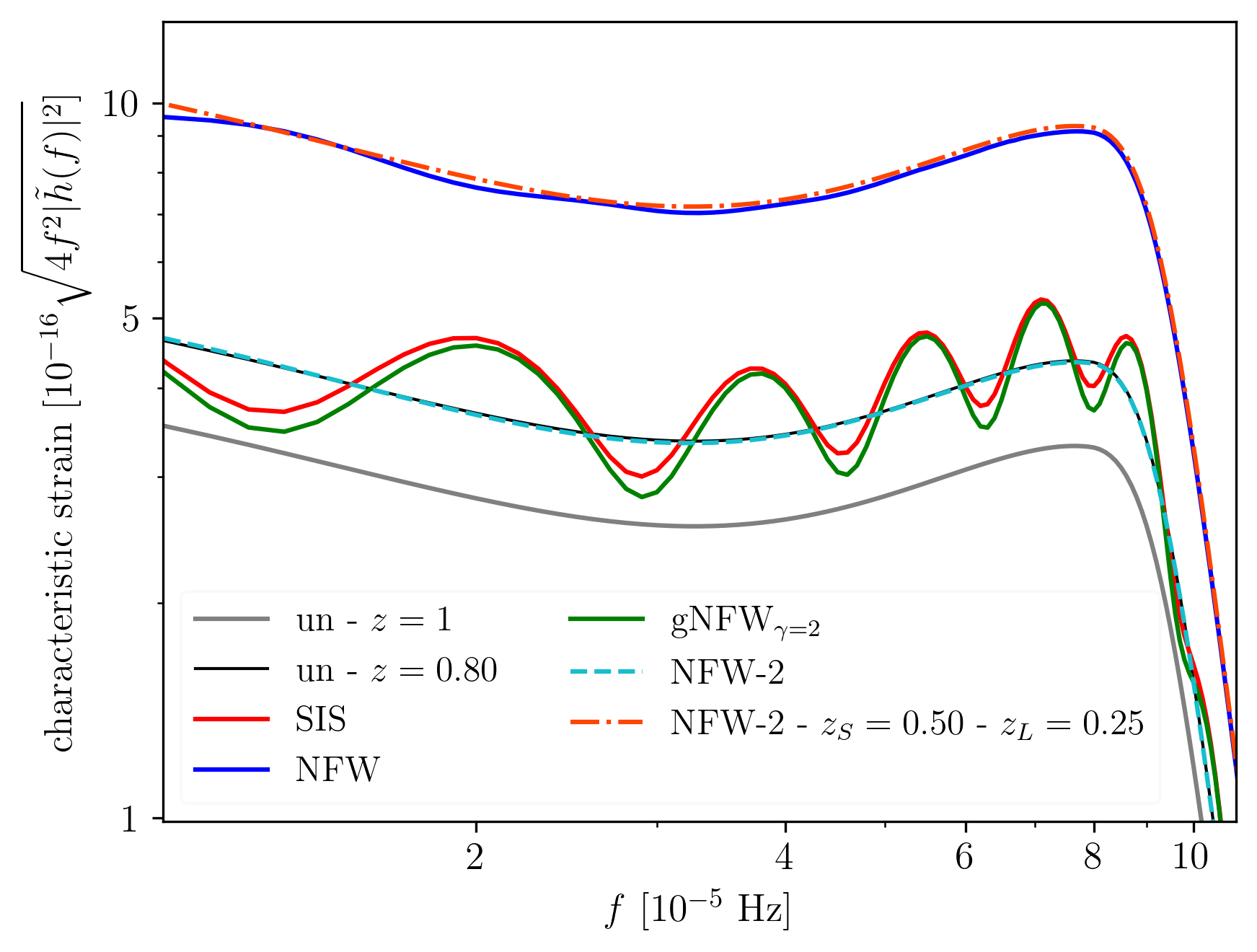}~~
    \includegraphics[width=0.48\textwidth]{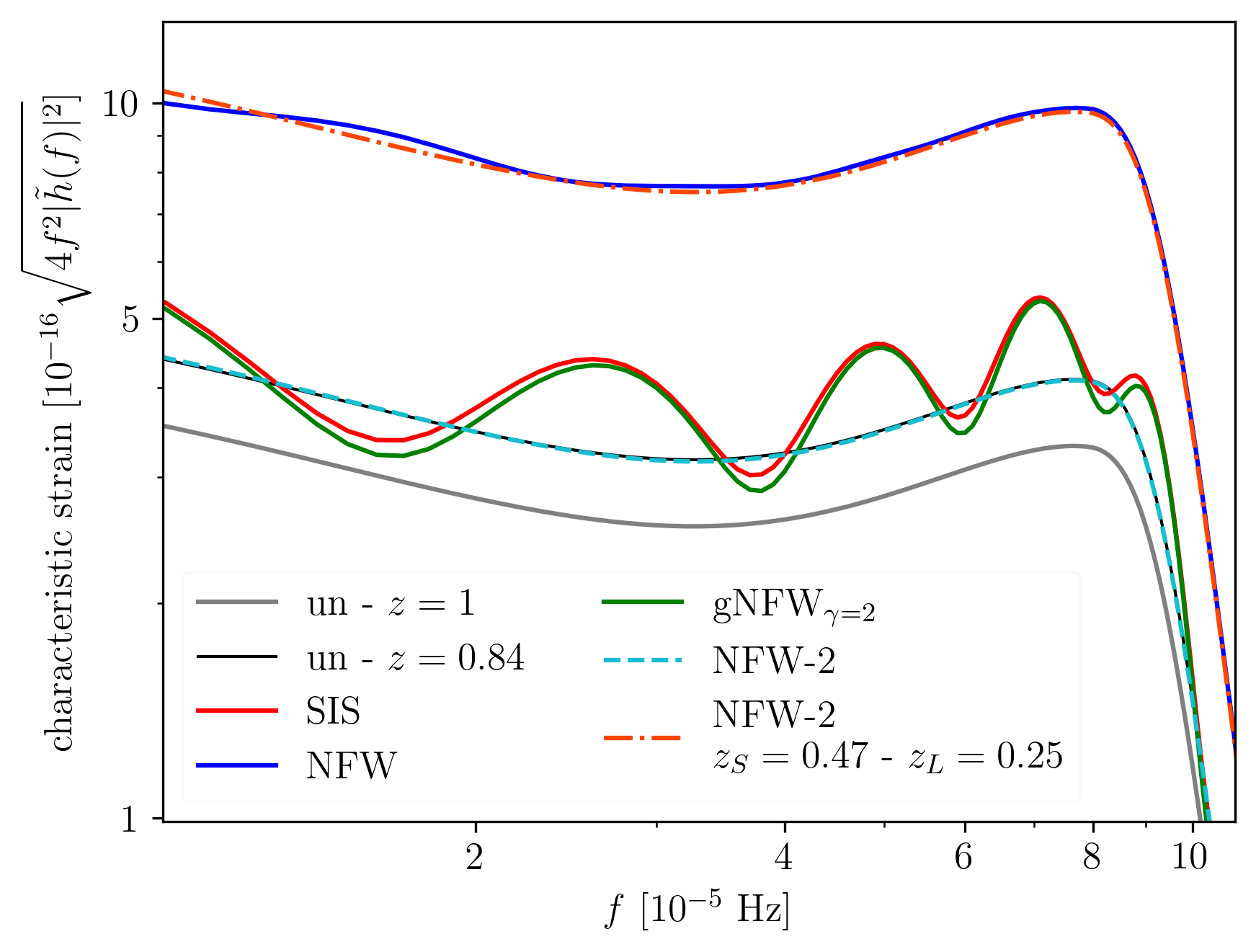}
    \caption{Frequency domain of different lensed waveform for the lens model that will be studied in the paper. Where it is not specified, the source is at $z_S=1$ and the lens is at $z_L=0.5$ in the \textit{left panel}, and at $z_L=0.15$ in the \textit{right panel}. All models have $y_{SIS}=1$, $M_L~=~10^9~M_\odot$, see text for more details. 
    }
    \label{fig:FDs}
\end{figure*}
  
\section{Phase effects} \label{ch:phase_effects}

One of the advantages of studying GWs with respect to EM signals lies in the way they are observed and in their sources. In fact, while an EM signal is usually monochromatic (the frequency of the signal, generally, does not change during an observation) and we can register only its frequency and magnitude, a GW one has a well-known time-frequency behaviour (dependent on the properties of the source) and we can ``register'' the very full waveform of the signal. On the other hand, though, in GW observations we lose in spatial resolution \cite{LIGOScientific:2018mvr,LISA2017}. 

\begin{figure*}[htbp]
    \centering
    \includegraphics[width=0.48\textwidth]{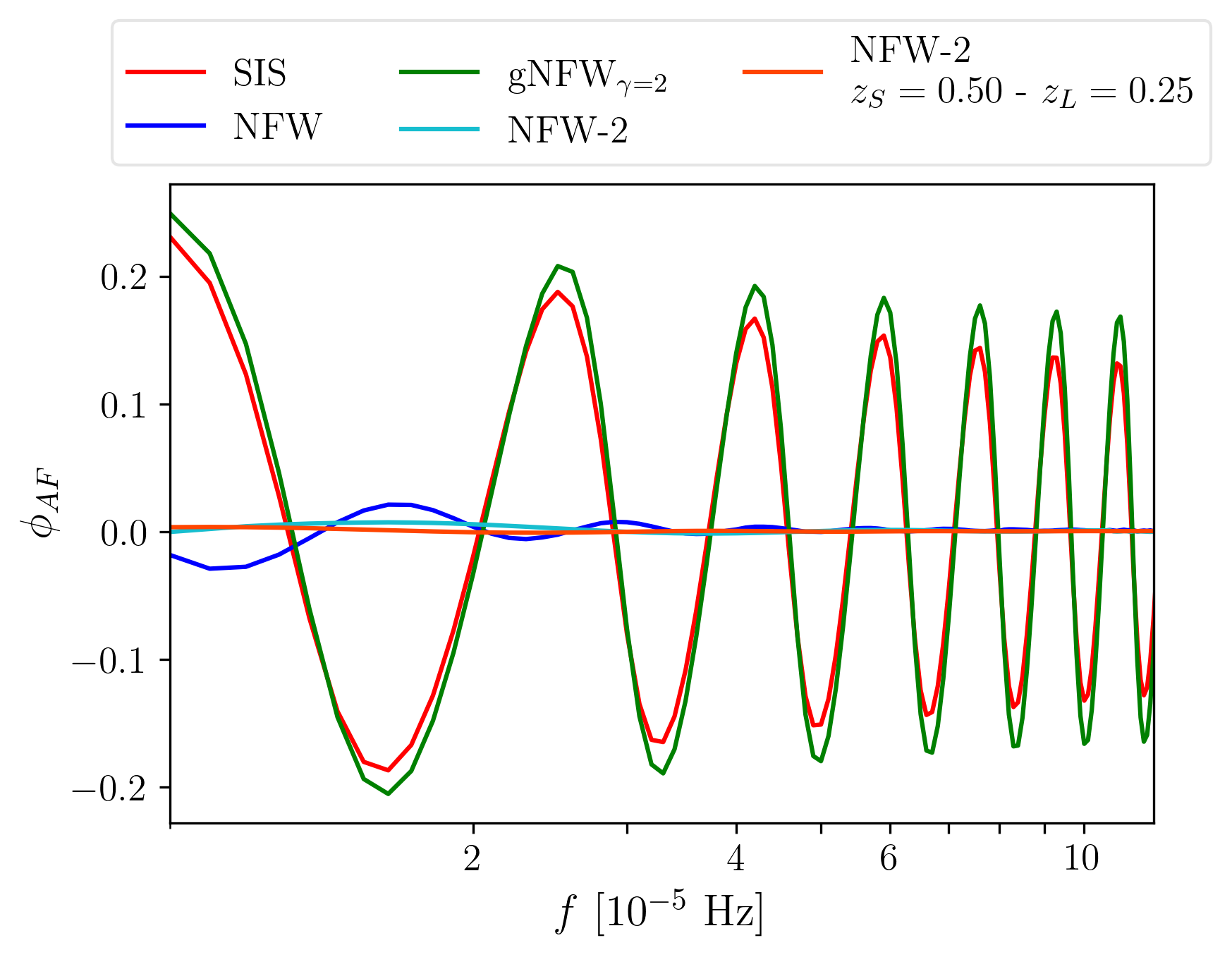}
    \includegraphics[width=0.48\textwidth]{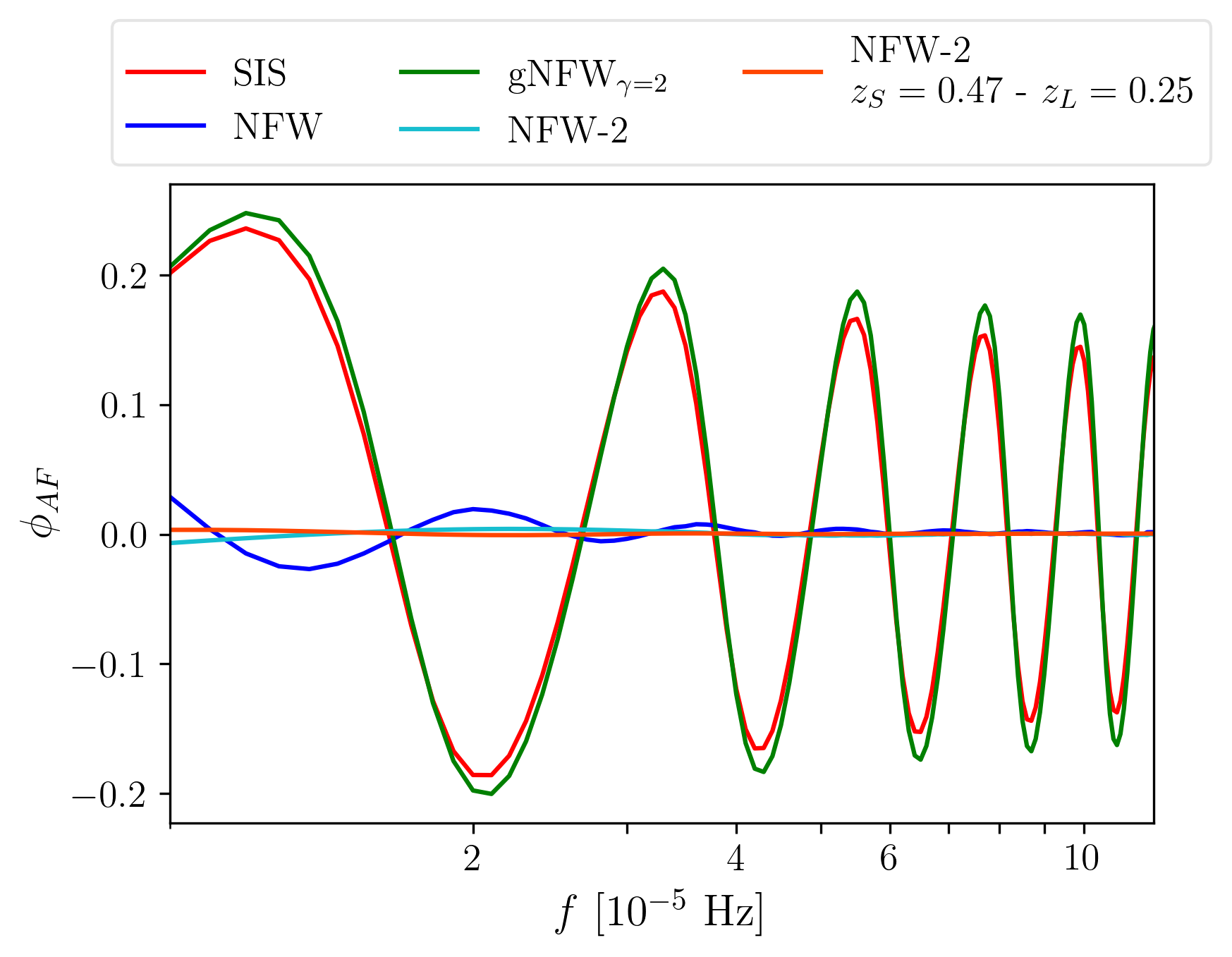}
    \caption{Phase of the amplification factor for the lens models taken into consideration in this study. \textit{Left panel}: lenses at redshift $z_L=0.5$; \textit{right panel}: lenses at redshift $z_L=0.15$. 
    }
    \label{fig:phases}
\end{figure*}

To have the possibility of studying the waveform includes the opportunity to take advantage of many types of information in it, like, for example, its phase. The phase of the unlensed signal (and its evolution) is set by the source configuration, in particular by the chirp mass\footnote{The chirp mass is defined as $\mathcal{M}=(m_1 m_2)^{3/5}/(m_1+m_2)^{1/5}$, where $m_1$ and $m_2$ are the masses of the black holes.} of the BBH \cite{Cutler:1994}. In the WO regime, this phase is modulated by gravitational lensing effects. In particular, the change in phase produced by the lensing can be defined from the phase of the AF as
\begin{equation}
    %\Delta 
    \phi_{AF} (w, y) = -i \log\left(\frac{F(w,y)}{|F(w,y)|}\right)~.
\label{eq:phase_shift}
\end{equation}
The phases of the AFs for the models taken into consideration are shown in Fig.~\ref{fig:phases}.
From the figure, we can see how different mass models imprint different phase shifts in the waveforms. In the next sections, we try to sort out if and how we can distinguish a lensed waveform from an unlensed one, in the different optics regimes defined by Eq.~\eqref{eq:GO_cond2}, by using only the phase of the signals.

\subsection{Wave Optics}

First, we consider the case of WO, where no interference patterns are present and the lensed waveform is modulated with respect to the unlensed one. Eq.~\eqref{eq:phase_shift} depends both on the properties of the source through $w$, which is a time dependent quantity, and on the characteristics of the lens model. We want to explore here if lens mass models may leave a unique imprinting on the waveform. 

For that purpose, as an example, looking to Fig.~\ref{fig:FDs}, if we had to judge only from the frequency (or time) domain waveforms, the lensed event with source at $z=1$ and a NFW-2 lens (light blue) at $z_L=0.5$ ($z_L=0.15$) would turn out to be identical to an unlensed waveform with source at $z=0.8$ ($z=0.84$) (black). But, since the phase shift from the AF acts on the lensed waveform, we could be able to use these phase effects for our goal.

The first tool we use to distinguish the two signals is the SNR ratio, in particular $\rho/\rho_{opt}$ (see Sec.~\ref{sec:SNR}), with the SNR calculated from Eq.~\eqref{eq:SNR}. In fact, the SNR is sensitive to the phase of the signal, since this information is stored in the frequency domain waveform, $\tilde{a}(f)$ and $\tilde{b}(f)$ in Eq.~\eqref{eq:inner_product}, that are used for the SNR calculation. In particular, the phase of the signal in the frequency domain can be calculated from the angles of the complex waveform functions (using, for example, the \texttt{numpy.angle} function in \textsc{Python}). 

Therefore, in the calculation of $\rho/\rho_{opt}$, where the template is given by the unlensed waveform and the signal by the lensed one, we would never get a value exactly equal to one. 
In this case, according to the parameter of the source and to LISA sensitivities \cite{Cornish:2018dyw}, the signal has $\rho\approx220$, very similar to the optimal one. In fact, we get $\rho/\rho_{opt}\approx1-4\cdot10^{-7}$. For such a signal this value has to be compared with a $3\sigma$ threshold of $\rho/\rho_{opt}\approx1-1.5\cdot10^{-4}$ that, as from Eq.~(\ref{eq:threshold}), means that if we had a signal with $\rho/\rho_{opt}<1-1.5\cdot10^{-4}$, it would be distinguishable from the template at $3\sigma$ level. It is clear that, in this case, a sole template match analysis cannot distinguish the unlensed signal from the lensed one. In fact, we would need a signal with $\rho \sim 4000$ to set the threshold at $\rho/\rho_{opt}\approx1-4\cdot10^{-7}$ and make the signals different at the given confidence level. Following the same line of reasoning, if we had a signal with $\rho=800$, that is still a realistic estimation for a LISA signal, we would have $\rho/\rho_{opt}~\approx~1-3\cdot10^{-5}$, that is inside the threshold and thus indistinguishable from the unlensed template. 
We can show, though, how it would be distinguishable from an unlensed one if phase effects were taken into account. 
\begin{figure*}[htbp]
    \centering
    \includegraphics[width=0.3\textwidth]{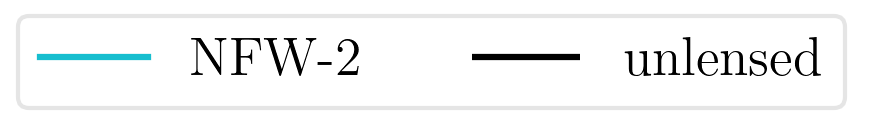}\\
    \includegraphics[width=0.48\textwidth, trim={0 0 0 1.2cm}, clip]{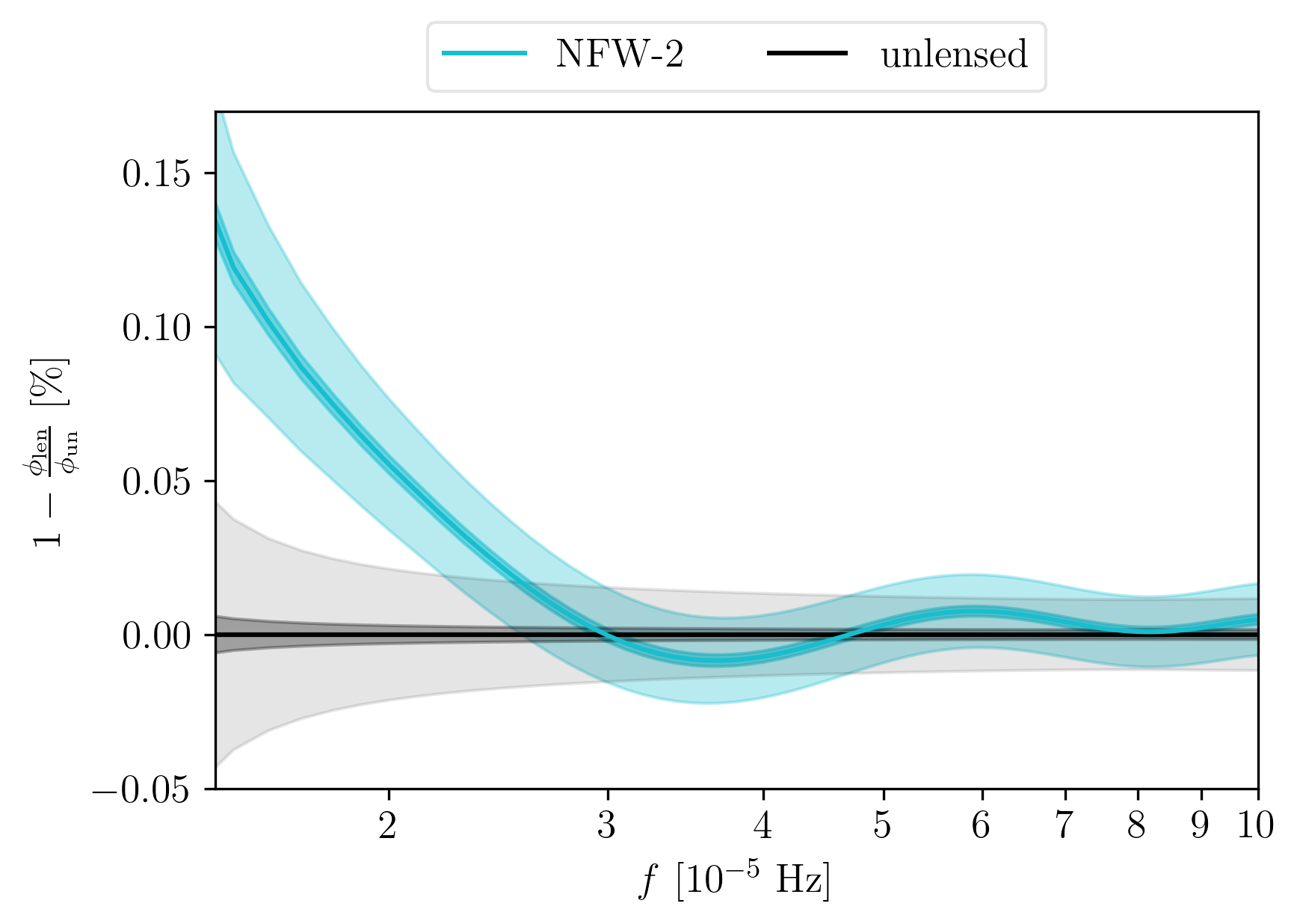}~~
    \includegraphics[width=0.48\textwidth, trim={0 0 0 1.2cm}, clip]{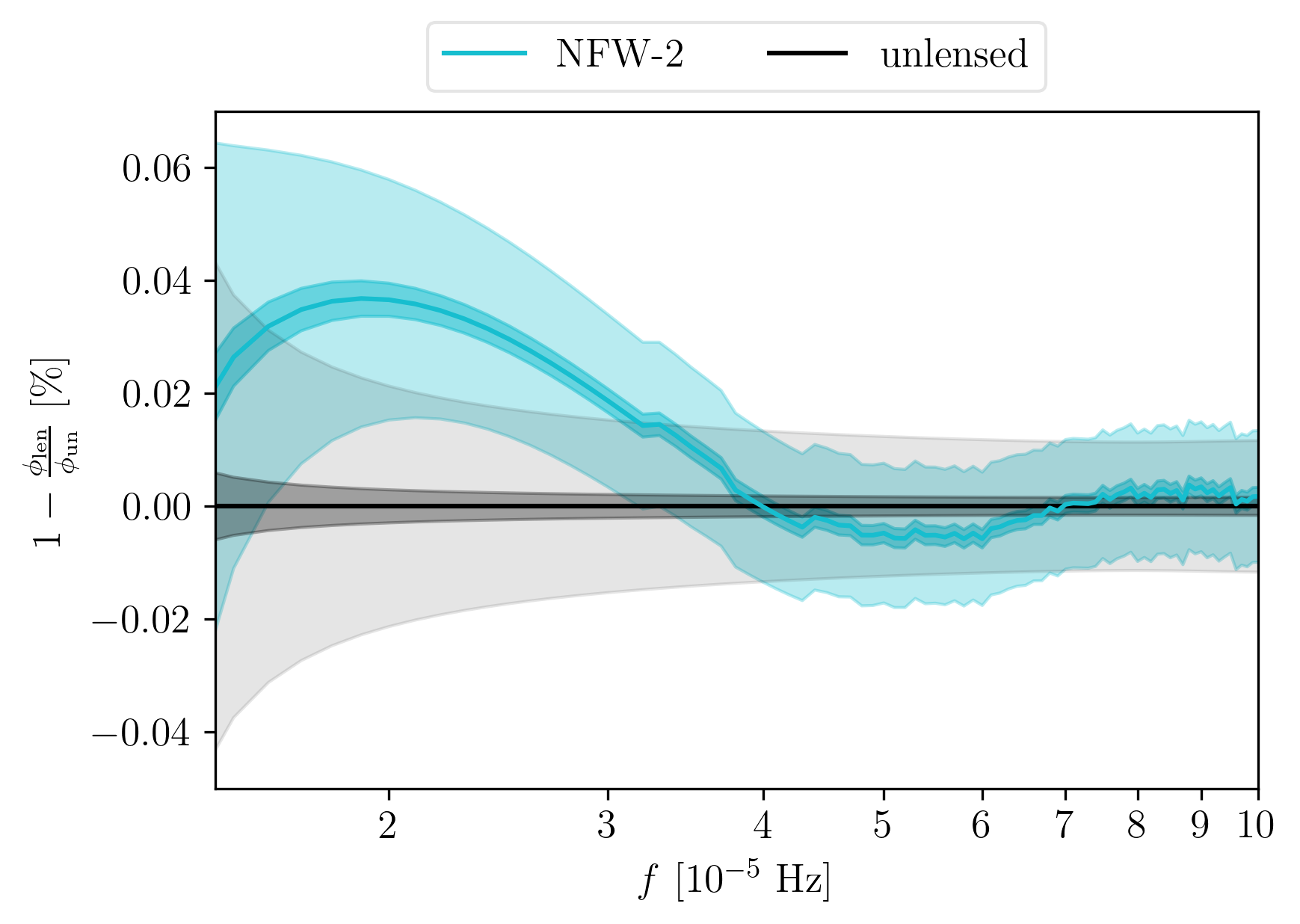}
    \caption{Percentual difference in the phase between lensed and unlensed signal. The error contours are calculated for the unlensed template (black) and lensed signal (blue) with SNR $\rho=220$ (lighter colors) and $\rho=800$ (darker colors). \textit{Left panel}: lenses at redshift $z_L=0.5$; \textit{right panel}: lenses at redshift $z_L=0.15$.
    % Inset: phase of the amplification factor for a NFW-2 lens at $z=0.5$.
    }
    \label{fig:NFW2_phases}
\end{figure*}

To this aim, we lead a study of the phases for a deeper analysis. In Fig.~\ref{fig:NFW2_phases}, we can see how the phases of the lensed (light blue) and unlensed (light black) waveform change with the frequency. In particular, we show the percentual difference between the lensed and unlensed phases.
The error associated with the phase is given by \cite{Cutler:1994} 
where it is shown that, in a matched filtering analysis, the phase of the waveform can be measured with an accuracy corresponding to the inverse SNR,
\begin{equation} \label{eq:err_phase}
    \sigma_\phi \approx \rho^{-1}~{\rm rad}~.
\end{equation}
As said before, the SNR for such a signal and LISA sensitivities is $\rho=220$. Since we are considering a normalized phase, the errors are computed propagating from the relatives one. From Fig.~\ref{fig:NFW2_phases} we can see how, at low frequencies ($f\lesssim3\cdot10^{-5}$), the unlensed waveform lies outside the confidence region of the lensed one, while the two are indistinguishable in the rest of the spectrum.

\begin{figure*}[htbp]
    \centering
    \includegraphics[width=0.48\textwidth]{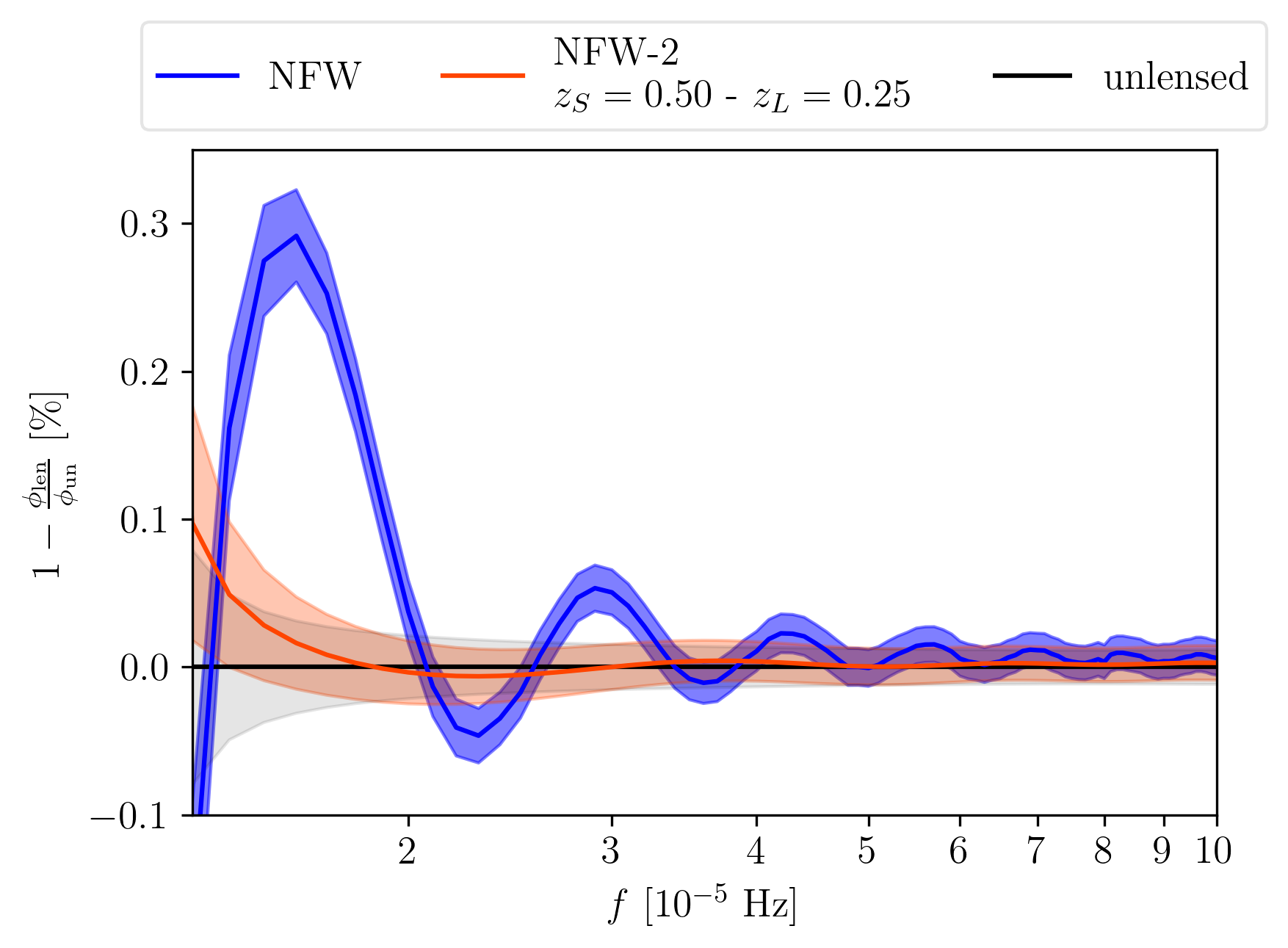}~~
    \includegraphics[width=0.48\textwidth]{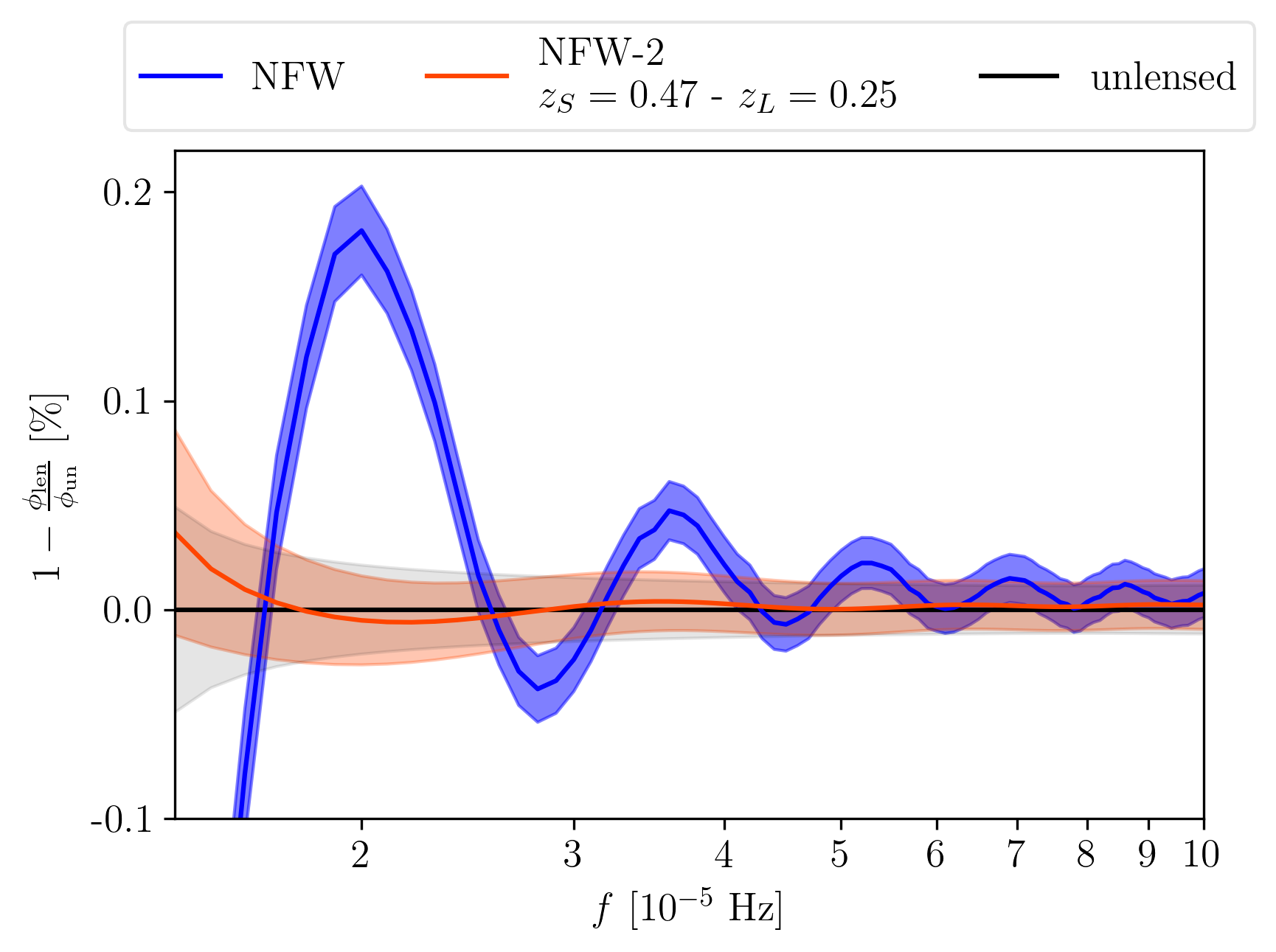}
    \caption{Percentual difference in the phase between different lensed signals and the unlensed one. In the \textit{left panel} (\textit{right panel}) we show comparison between NFW (blue) and NFW-2 (orange) lens model, at $z_L=0.5$ ($z_L=0.15$) and $0.25$, respectively, described above in the text. The source for NFW is at $z_S=1$, while the source of NFW-2 is at $z_S=0.5$ ($z_S=0.47$). 
    The error bars in both panels are computed for the signal SNR, $\rho=220$. 
    }
    \label{fig:NFWvsNFW2}
\end{figure*}

A similar qualitative analysis can be seen for the cases shown in Fig.~\ref{fig:NFWvsNFW2}. There, the NFW-2 lens produces phase effects that are not clearly distinguishable from the unlensed signal, given that the SNR in this case of $\rho=220$. On the other hand, stronger effects can be seen from the NFW lens, which is quite different from the unlensed case in the lower range of frequencies.

Can such a study on the phase provide us with further (than only template matching) new and crucial elements to help us to distinguish the two signals and thus detect to a higher confidence a lensed signal? The answer strongly depends on both the magnitude of the effect and of the signal because, as shown in Eq.~\eqref{eq:err_phase}, the error on the phase depends on the SNR of the signal. Thus, while with a signal with $\rho=220$ we have no clear distinction between the unlensed and the lensed case, the situation improves at higher SNR. As an example, in Fig.~\ref{fig:NFW2_phases}, we plot the error for a signal with $\rho=800$ (dark blue), that is still a realistic estimation for a LISA signal. We can see that, for such a signal, even though the $\rho/\rho_{opt}$ value is still inside the threshold, the phases of the waveforms can be distinguished almost throughout the whole spectrum of the signal.

\subsection{Interference regime
}\label{ch:int}

We move now to the interference regime, that stands in between the GO and WO ones and is defined by the condition $f\cdot \Delta t \sim 1$, where $f$ is the frequency of the lensed signal and $\Delta t$ the time delay between the lensed images. This regime is characterised by interference patterns in the lensed waveform. 

\begin{figure*}[htbp]
    \centering
    \includegraphics[width=0.35\textwidth]{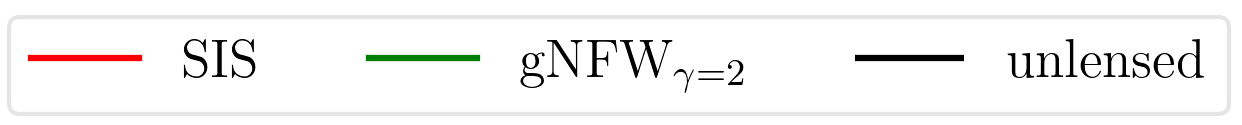}\\
    \includegraphics[width=0.48\textwidth, trim={0 0 0 1.2cm}, clip]{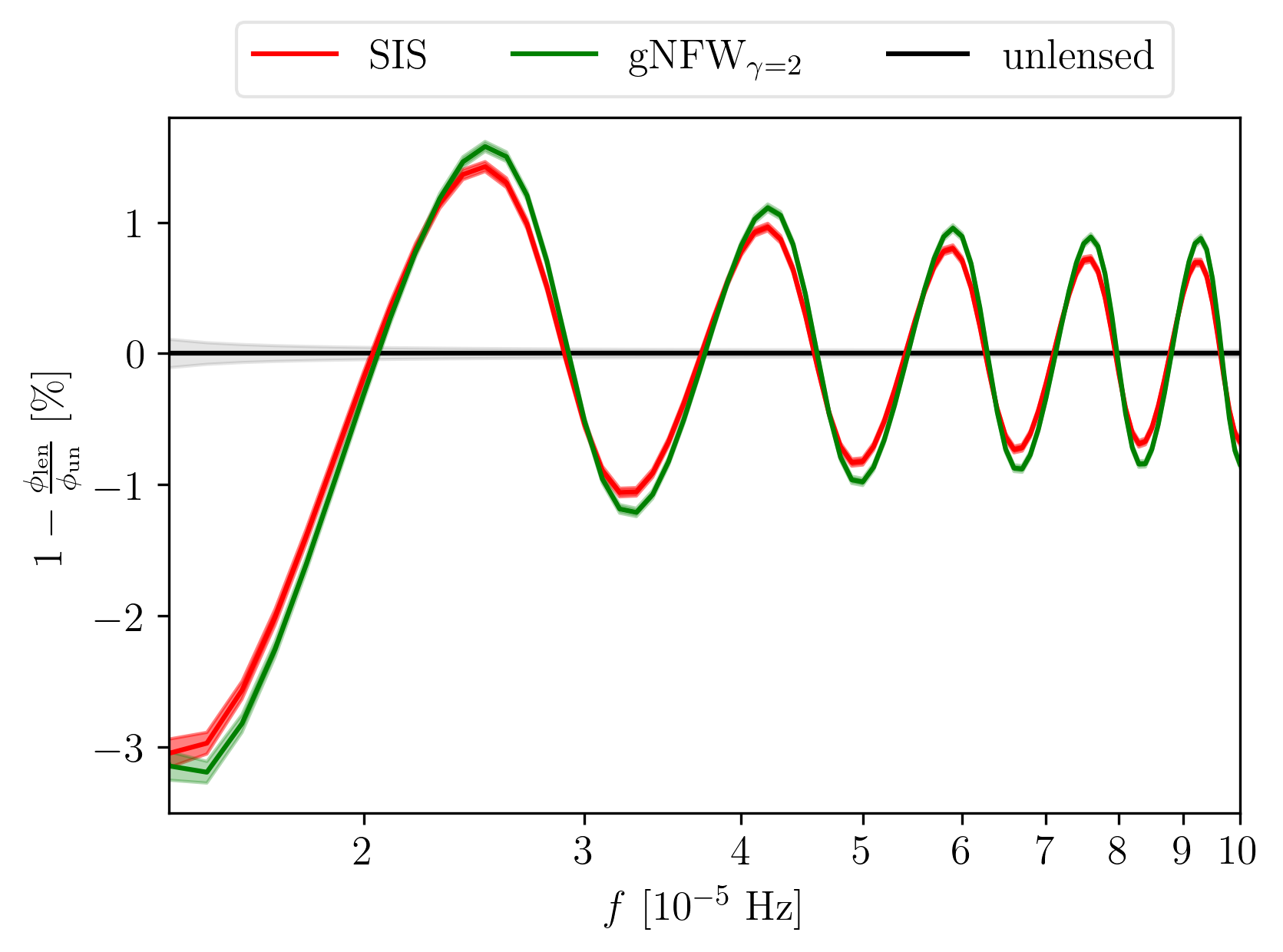}%\\
    \includegraphics[width=0.48\textwidth, trim={0 0 0 1.2cm}, clip]{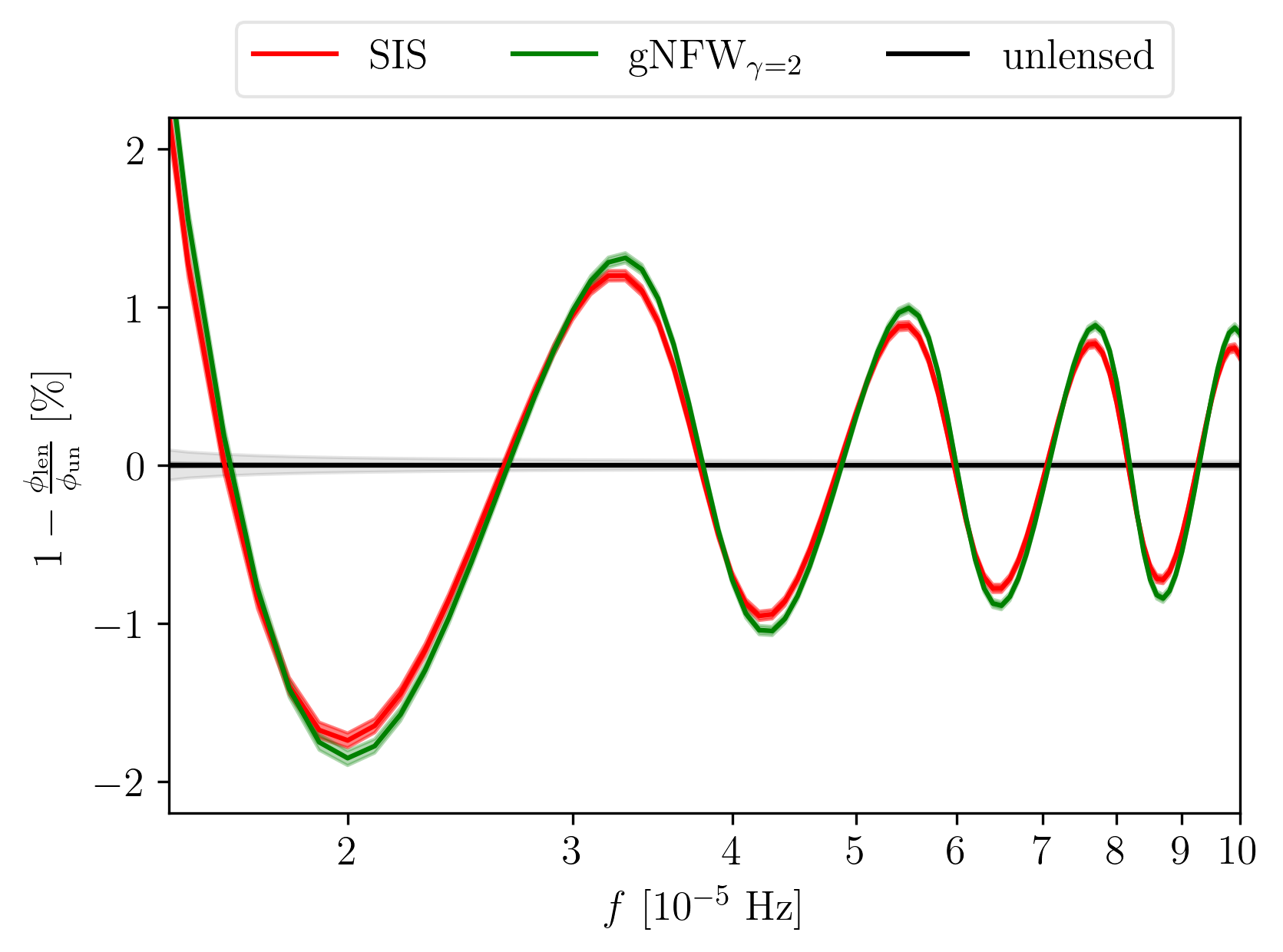}
    \caption{Percentual difference on the phase of different lensed signals and the unlensed one. Comparison between SIS (red) and gNFW$_{\gamma=2}$ (green) lens model, described above in the text. 
    \textit{Left panel}: lenses at redshift $z_L=0.5$; \textit{right panel}: lenses at redshift $z_L=0.15$.
    The error bars in both panels are given by a signal with $\rho=100$. 
    }
    \label{fig:SISvsGNFW2}
\end{figure*}

For our study on distinguishing lensed from unlensed signal, here we compare the SIS case with the unlensed one. In this case, the phase effects are clearly evident. In Fig.~\ref{fig:FDs}, the red waveform depicts the case of a SIS lens: although it has the same mass and physical impact parameter of, for example, the NFW-2 case described above (light blue waveform) it is clear how they have a much different impact on the waveform. Additionally, in Fig.~\ref{fig:SISvsGNFW2} we show the phase of the signal w.r.t. the frequency and we compare it with the phase of the unlensed one (in black). Now the error region is given by the SNR of the signal, $\rho=100$. We can see how the oscillatory behaviour of the lensed phase and the magnitude of the lensing effect ease the distinction between the two signals.
The phase of the AF for the SIS model is shown in Fig.~\ref{fig:phases}, in red. As we can see, the effects are much larger than the previous NFW-2 case (light blue curve in Fig.~\ref{fig:phases}), which explains why it is so easily recognizable.

The SNR calculations confirm what was just stated. If we consider, again, the unlensed waveform as template, and the lensed one as signal, we get $\rho/\rho_{opt}=0.11$. The threshold at $3\sigma$ for a $\rho=100$ is $\rho/\rho_{opt}=1-6\cdot10^{-4}$, which means that the template and the signal are clearly distinguishable. Even for lower SNR values, the two waveforms can be easily told apart (for $\rho=10$ we have a threshold of $\rho/\rho_{opt}=0.941$), but realistic values from LISA detector of SNR are much higher, $\sim 10^2 - 10^3$ \cite{Robson:2018ifk}.

The same conclusions can be derived when focusing on the gNFW$_{\gamma=2}$ model, depicted by the green curves in the plots of Fig.~\ref{fig:SISvsGNFW2}. The only (tiny) difference is given by the fact that, since this model has three free parameters, the threshold is given by $\rho/\rho_{opt}=1-7\cdot10^{-4}$. 

\subsection{Geometrical Optics}

We are not particularly interested in this regime since there is a thorough study in \cite{Ezquiaga:2020gdt} on what is the role of phase effects in GO, and how it can be used to recognize lensed events. 

\section{Lens Models Effects} \label{ch:models}

One important point to be discussed now is: once we recognize a signal as lensed, can we constrain the lens mass model? This is a crucial point to investigate, since a precise modelling of the lens can lead to further relevant studies, e.g. in the dark sector. 
In this section, we clearly show that such a constrain on the model is feasible in most of the cases and, even when the lensed waveforms are almost identical (as, for example, it is the case of SIS and gNFW$_{\gamma=2}$), we can still rely on additional elements of the signal to distinguish among them.

Once again, let us remind that in Fig.~\ref{fig:FDs} we show the scenario we are going to study: we have different lens models that produce different patterns in the lensed waveform, even if they have similar mass profiles, as shown in Fig.~\ref{fig:masses}, and the same dimensionful impact parameter. From the figure, it looks like the NFW and NFW-2 lenses only produce a magnification of the unlensed signal (or, at most, tiny oscillations w.r.t. the unlensed case). In the previous section, we have shown how to deal with this problem.

The SIS model, on the other hand, imprints some characteristic features visible in the lensed waveform that make it easy, along with the different magnification, to identify the model. The gNFW$_{\gamma=2}$, though, also presents features that are very similar to the SIS one.

\subsection{SIS vs NFW} \label{ch:SISvsNFW}

Looking at Fig.~\ref{fig:FDs}, one can see that to differentiate the NFW model (blue) and the SIS or gNFW$_{\gamma=2}$ ones (red and green, respectively) should be straightforward since the lensed waveforms are very different, both in magnitude and intrinsic features.

There is not even the need to study the phases, in this case. In fact, if we consider the SIS waveform as template and the NFW one as signal, we find $\rho/\rho_{opt}=0.56$, with a $\rho_{opt}\sim100$. Since the $3\sigma$ threshold for such a SNR is $\rho/\rho_{opt}~\approx~0.9994$, it is impossible that one could misinterpret the SIS case with the NFW one.

At this point, a much more difficult question to answer is whether a different combination of parameters in the SIS lens system could give place to a lensed waveform identical to the NFW one, or vice-versa. In principle, if one could change both source and lens parameters, the answer is yes. If two different sets of parameters ``fit" the signal (with the same precision) there is little one can do, apart from searching for further independent data (e.g., galaxies catalogs, EM data of the lens, etc$\ldots$). 

Something important to be checked now is up to which limits/ranges the two models would be indistinguishable. This problem, though, is of complex quantification since we would need to run an inference analysis on our NFW and SIS signals, which unfortunately overcomes the computational power at our disposal. That is because, apart from the usual inference analysis to be performed on the source, which already involves many templates and corresponding parameters, one should also add the inference part connected to the lens system, making the whole problem very heavy on the computational side. 
Nevertheless, taking advantage of calculations from Sec.~\ref{ch:phase_effects} and below, in Sec.~\ref{ch:conclusions} we try to give a more quantitative estimate on the conditions that make two signals distinguishable.

\subsection{SIS vs gNFW} \label{ch:SISvsGNFW2}

Here, we discuss whether we can distinguish the case involving the SIS lens (in red in Fig.~\ref{fig:FDs}) from the case with gNFW$_{\gamma=2}$ (in green). Again, we compare models with the ``same" projected mass and choose the same dimensionful impact parameter (that makes $y$ different for each of them).

In this case, the SNR comparison between the two gives $\rho/\rho_{opt}=0.9868$ with $\rho_{opt}=98.9355$, where we considered the waveform lensed by the SIS model as the template and the gNFW$_{\gamma=2}$ as the signal. In principle, the two waveforms are distinguishable given the SNR of the signal. In fact, the threshold for such a signal is set at $\rho/\rho_{opt}~\approx~0.9994$, meaning that signals with lower values of $\rho/\rho_{opt}$ are distinguishable from the template. Moreover, as we showed in Sec.~\ref{ch:int}, and considering such a threshold, they should be easily distinguishable from an unlensed signal, as well. In fact, $\rho/\rho_{opt}=0.1305$ if we consider as template the unlensed waveform and the lensed signal as with gNFW$_{\gamma=2}$, and $\rho/\rho_{opt}=0.1056$ when the signal is a SIS lens. 
The same happens if we have a closer unlensed signal as template, since SNR calculations are not sensitive to the magnitude of the strain of the waveform, as one can see from Eq.~\eqref{eq:SNR}.

\begin{figure*}[htbp]
    \centering
    \includegraphics[width=.25\textwidth]{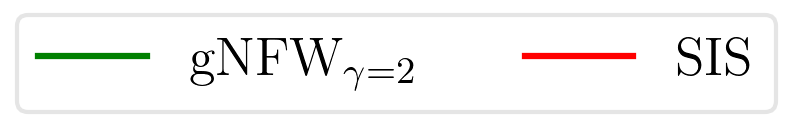}\\
    \includegraphics[width=0.48\textwidth, trim={0 0 0 1.2cm}, clip]{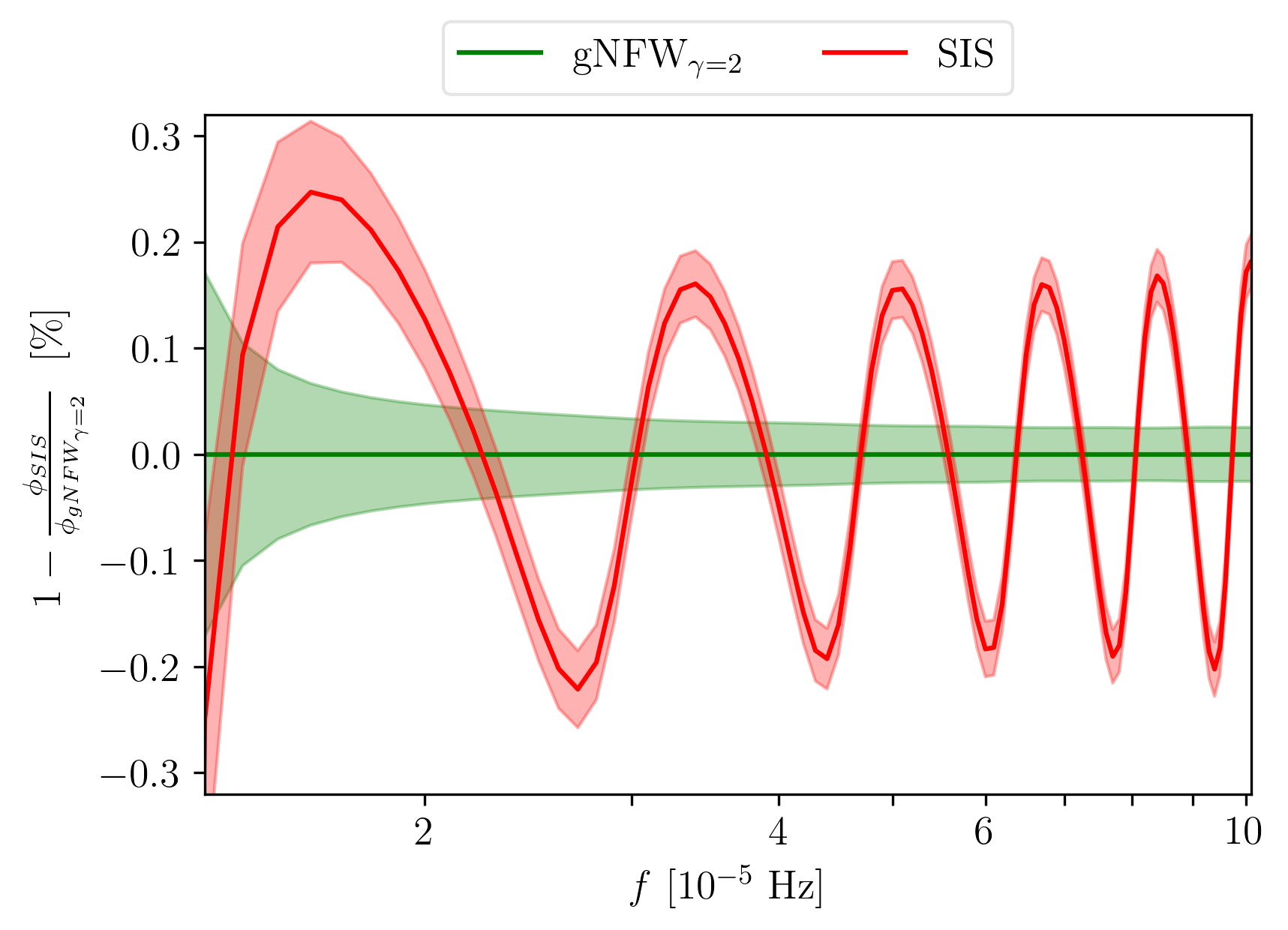}
    \includegraphics[width=0.48\textwidth, trim={0 0 0 1.2cm}, clip]{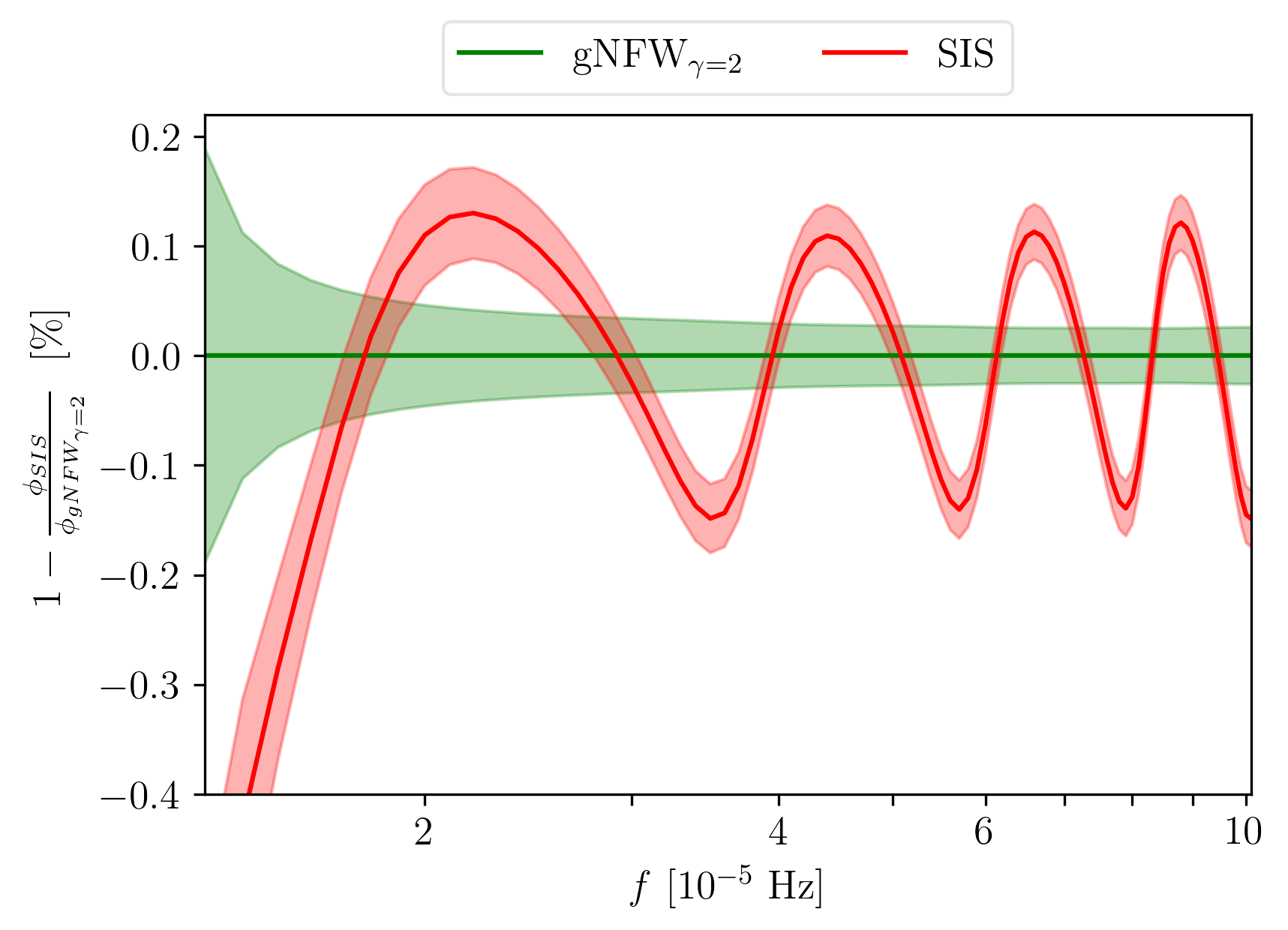}
    \caption{Percentual difference on the phase of two different lensed signals. Comparison between SIS (red) and gNFW$_{\gamma=2}$ (green) lens model, described above in the text. 
    \textit{Left panel}: lenses at redshift $z_L=0.5$; \textit{right panel}: lenses at redshift $z_L=0.15$. The error bars in both panels are given by a signal with $\rho=100$. 
    }
    \label{fig:SISvsGNFW2_PE}
\end{figure*}

The same conclusions can be extracted by looking at Fig.~\ref{fig:SISvsGNFW2_PE}. 
There, we show the lensed phases normalized at the gNFW$_{\gamma=2}$ one. The errors on the phase are given by Eq.~\eqref{eq:err_phase}, for a signal with $\rho=100$, that is the SNR of this event for LISA sensitivities.
The lensing effects give the phases the oscillatory behaviour that can be seen on the plot, and that makes it easy to distinguish not only the lensed signal from the unlesed one, but also between the two different models. In fact, even though the lensed phases have basically the same behaviour, we can set them apart because their errors are relatively small when compared to the previous cases and because of the magnitude of the lensing effect. 

\subsection{NFW vs NFW-2} \label{ch:NFWvsNFW2}

This case is of particular interest. Fig.~\ref{fig:FDs} shows the comparison between a waveform given by the lensing effect of a NFW lens at $z_L=0.5$ ($z_L=0.15$ in the right panel) on a signal from a source at $z_{S}=1$ (blue), and one given by a NFW-2 lens at $z_{L}=0.25$ and a source at $z_{S}=0.5$ ($z_S=0.47$) (orange dot-dashed). First of all, as usual, we try to match the two waveforms, to see if they are distinguishable through the SNR method. As we can see from Fig.~\ref{fig:FDs}, the two waveforms are identical, and this is confirmed by the value of the match, i.e. $\rho/\rho_{opt}=1-1.4\cdot10^{-6}$. In the calculation, we considered the NFW system as the template and the NFW-2 one as the signal. Given a signal with $\rho\approx220$, the threshold at $3\sigma$ for three free parameters is $\rho/\rho_{opt}~\approx~1-1.4\cdot10^{-4}$, meaning that we are far from being able to tell the two signals apart. We would need a signal ten times stronger, i.e. with $\rho=2200$, to reach the needed precision. 

\begin{figure*}[htbp]
    \centering
    \includegraphics[width=0.48\textwidth]{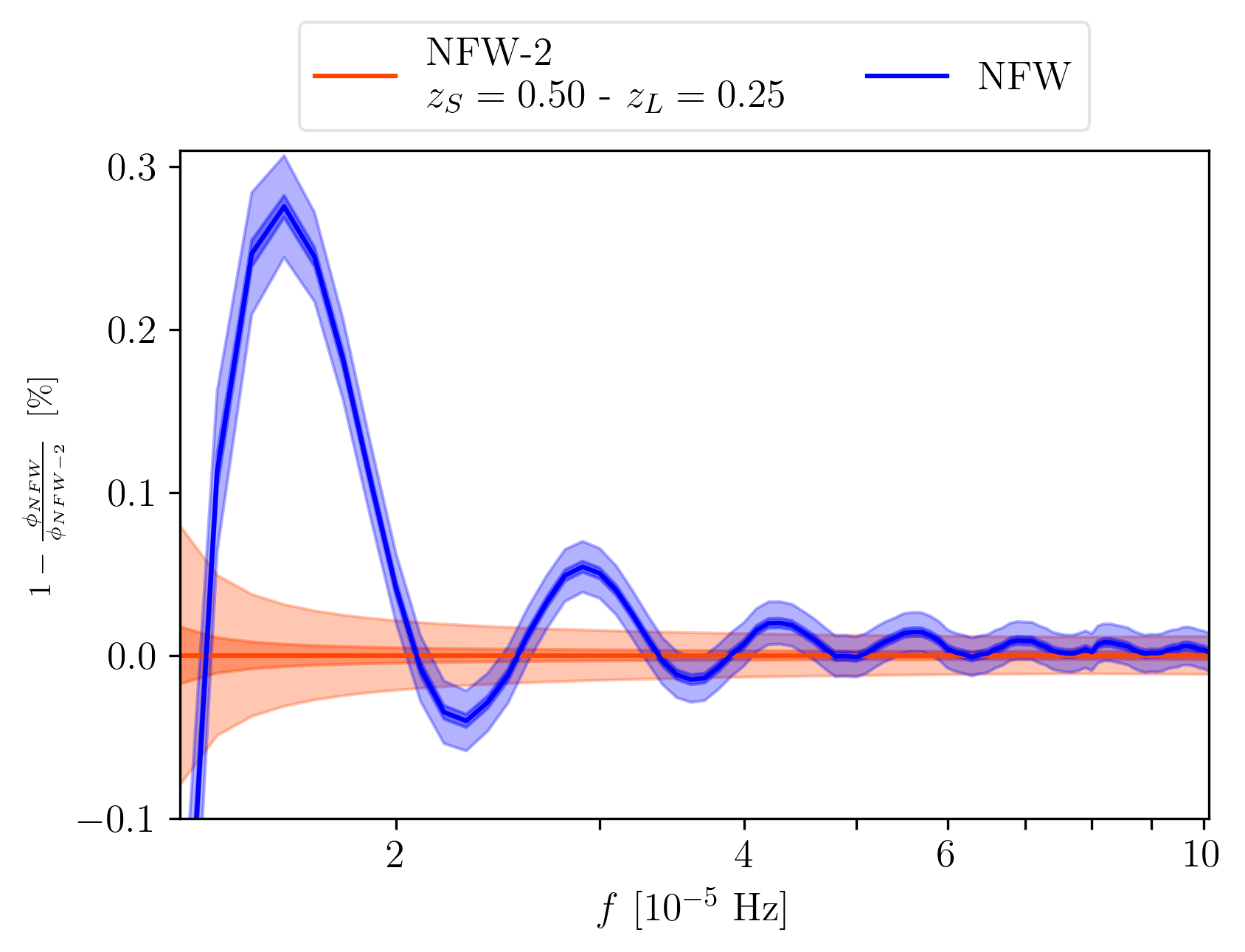}
    \includegraphics[width=0.48\textwidth]{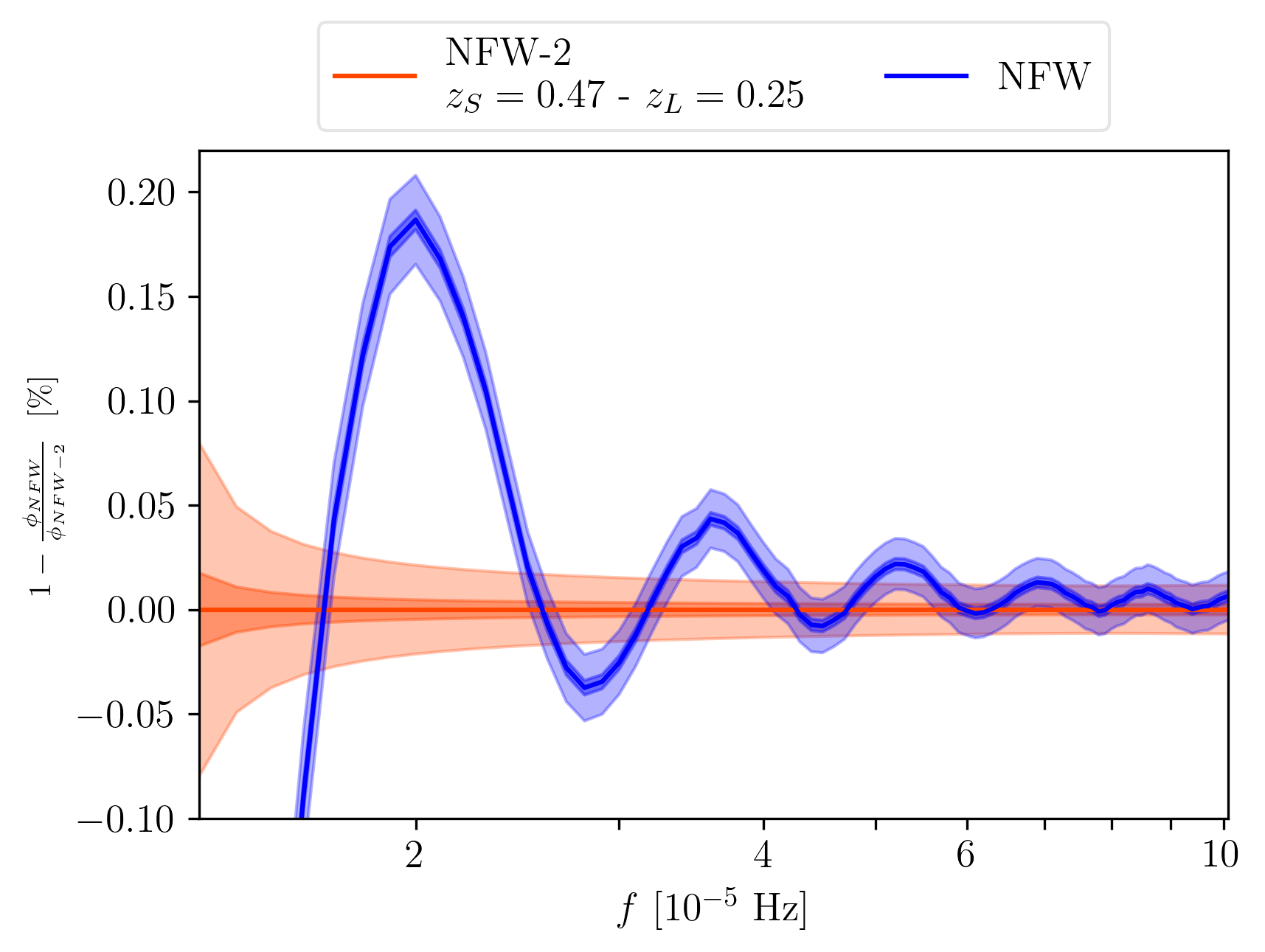}
    \caption{Percentual difference on the phase of two different lensed signals. In the \textit{left panel} (\textit{right panel}), we see the comparison between NFW (blue) and NFW-2 (orange) lens model, at $z_L=0.5$ ($z_L=0.15$) and $0.25$, respectively, described above in the text. The source for NFW is at $z=1$, while the source of NFW-2 is at $z_S=0.5$ ($z_S=0.47$). The error bars are computed for a signal with $\rho=220$. The smaller confidence region are given by $\rho=1000$. 
    }
    \label{fig:NFWvsNFW2_PE}
\end{figure*}

As before, though, studying the phases of the waveforms comes in handy. In Fig.~\ref{fig:NFWvsNFW2_PE}, we show the phases of the lensed waveforms for the cases explained above. In the figure, we normalize the phases at the NFW-2 one. The errors on the absolute phases are given by Eq.~\eqref{eq:err_phase} for a signal with $\rho=220$, while the darker confidence regions are given for a source with $\rho=1000$. We chose to add the error for such a SNR because it is the minimum value for which the two signals are distinguishable.

We can see how, while the phases are different, in particular NFW has a larger magnitude, the two signals are indistinguishable for frequencies $f\gtrsim3\cdot10^{-5}$. As we said above, this is due both to the fact that the impact of the phase effect is low and the magnitude of signal is not big enough. In fact, if the signal has $\rho=1000$ instead of $\rho=220$, we would be able to distinguish the two signals, as shown in Fig.~\ref{fig:NFWvsNFW2_PE}. In particular, considering the darker regions, which are the confidence intervals given by $\rho=1000$, we can see how the 2 signals can be distinguished. This could be possible despite the SNR being smaller than the one required by the SNR method, i.e. $\rho=2200$.

\section{Inference} \label{sch:inference}

Once we are sure about the model, how well can we constrain the lens parameters? In our previous work \cite{Cremonese:2021puh}, we showed that in the interference regime the mass-sheet degeneracy can be broken. There, we considered a PM lens model, so it is natural to wonder whether the results are still valid for extended mass models. 

Considering a SIS lens, we make different calculations, summarized in Tab.~\ref{tab:SISmasses_ys}. There, we compare the SIS lens with $M_L=10^9$~M$_\odot$ and $y=1$, with the same model but different masses and $y$, and we show the value of $\rho/\rho_{opt}$. This is done considering the waveform with lens mass $M_L=10^9$~M$_\odot$ and $y=1$ as the signal, and the one with different mass and $y$ as the template. From Eq.~\eqref{eq:threshold}, 
\begin{equation}
    \frac{\rho}{\rho_{opt}}=1-\left[\frac{1}{2}\frac{\Delta\chi^2}{\rho_{opt}^2}\right]~,    
\end{equation} 
with $\Delta\chi^2\approx11.8$, 
and given that the SNR of the signal is $\rho\approx100$, the threshold for which two signals can be distinguished at $3\sigma$ level is $\rho/\rho_{opt}\approx0.9994$. Thus, the results of the table show how the breaking of the MSD is still valid, since we can distinguish between the different cases. For more details, see \cite{Cremonese:2021puh}. 
\setlength\extrarowheight{5pt} 
\renewcommand{\arraystretch}{1.2}
\begin{table}[htbp]
\begin{tabular}{|c|c|c|c|c|c|}
\dtoprule 
\diagbox{M$_L$\\ $[M_\odot]$}{$y$} & 0.8 & 0.9 & 1 & 1.1 & 1.2 \\ \hline
$10^8$               &  $0.9660$   & $0.9598$ & $0.9597$  & $0.9655$ & $0.9735$ \\ \hline
$2.5\cdot10^8$       & $0.9501$ & $0.9688$ & $0.9752$  & $0.9744$ & $0.9760$ \\ \hline
$5\cdot10^8$         & $0.9521$ & $0.9466$ & $0.9676$ & $0.9775$ & $0.9839$\\ \hline
$7.5\cdot10^8$       & $0.9566$ & $0.9561$ & $0.9755$ & $0.9821$ & $0.9720$ \\ \hline
$10^9$               & $0.9723$ & $0.9316$ & $1$   & $0.9629$ & $0.9897$ \\ \hline
$2.5\cdot10^9$       & $0.9204$ & $0.9558$ & $0.9738$ & $0.9814$ & $0.9844$ \\ \hline
$5\cdot10^9$         & $0.8680$ & $0.9375$ & $0.9715$ & $0.9822$ & $0.9852$ \\ \dbottomrule
\end{tabular}
\caption{Comparison in terms of $\rho/\rho_{opt}$ between different $M_L$ and $y$, considered as templates, with $M_L=10^9$ and $y=1$, considered as signal, for SIS lens.}
\label{tab:SISmasses_ys}
\end{table}

For the NFW case (pure WO regime), relatively small changes of $y$ or $M_L$ would only change the magnification, so there would be degeneracy. If the phase effect is too small to be seen, then, a pure magnification effect is degenerate with the source distance (and could be with other lens parameters, too). Thus, if we only have one image with magnification effects only, this could be misinterpreted with a closer unlensed signal or by a lens with higher mass/lower $y$.
As said above, though, if the phase effects are big enough (see e.g. Sec.~\ref{ch:NFWvsNFW2}) and/or the signal is bright enough, then, we can use these effects to break the degeneracy, since the phase would constrain the lens parameters. 

\section{Conclusions} \label{ch:conclusions}

In this study, in the context of a LISA signal characterized by gravitational lensing in the wave-optics regime, we had two goals: $i)$ to show how one can distinguish a lensed signal from an unlensed one; $ii)$ to study how different lens models leave different imprints on the lensed signal so that we can distinguish and constrain them. To achieve these goals we used two different tools: matching templates with signal-to-noise ratio calculation and phase effects.

The first method is given by a matched analysis, i.e. to calculate $\rho/\rho_{opt}$ comparing the signal with a chosen template. 
When we are not in the geometrical-optics regime, i.e. where we cannot apply the stationary phase approximation anymore, the phase shift caused by the lensing changes with the frequency and, then, it leaves a unique imprint in the phase of the unlensed signal. The SNR being sensitive to the phase of the signal, we would always be able to detect a lensed event. In reality, when the lensing effect is small and/or the magnitude of the signal is (relatively) low, we can not distinguish a lensed from an unlensed signal. Therefore, even though this method is useful, we showed that it is not always effective.

The second method consists on studying the phase of the signal. Investigating the phase effects is a powerful tool. In fact, as for the SNR method, we are always able to look at the phase and distinguish a lensed from an unlensed signal. Again, in a real case scenario, when the lensing effect is small so can be the phase shift, and it could result very difficult to detect a real signal (e.g. see Fig.~\ref{fig:NFW2_phases}). Nonetheless, for bright enough signals, or strong enough lensing effects, this is an effective way to find out if an event was lensed or not. Moreover, we showed how this method is more effective than the template matching/SNR one.

It also comes in handy when we have to distinguish between different lens models that give two (almost) identical lensed waveform. As we showed in Fig.~\ref{fig:NFWvsNFW2}, even though two lens models give the same lensed waveform, they could imprint different phase shifts on the signal, i.e. the imprints can be differentiated. Again, the effects and/or the signal have to be high enough. In this case, as well as before, studying the phases is more effective than simply considering the template matching. 

For example, when comparing the NFW and the NFW-2 models as in Fig.~\ref{fig:NFWvsNFW2_PE}, a SNR of $\rho=2200$ is needed to distinguish the two signals using matched filtering analysis only. On the contrary, by exploiting the phase effects, the signals need to have a SNR of $\rho\approx800$, almost one third of the SNR needed in matched analysis.

When the lensed waveform presents characteristic features (e.g. Fig.~\ref{fig:SISvsGNFW2}), recognition and inference become easier. Even when we have two models that look (almost) identical, we can tell them apart, as we showed for SIS and gNFW$_{\gamma=2}$ in Sec.~\ref{ch:SISvsGNFW2}.
More precisely, referring to Fig.~\ref{fig:phases}, when the wave effects are large, i.e. $\phi_{AF}\geq0.1$, there is no problem to distinguish a lensed waveform from an unlensed one and to characterize correctly the lens model, even for SNR as low as $\rho\approx100$. On the other hand, when $\phi_{AF}<0.1$, we need signals with high SNR, of the order of $\rho\approx2\cdot10^3-4\cdot10^3$ when using only matched filtering analysis. The situation improves when considering the phase of the signals: in this case, in fact, we need signals with $\rho\approx800-1000$.

This study can be of great importance when investigating the dark sector of the universe. In fact, a precise modelling of the lens can help in constraining the dark matter component in galaxies \cite{Diego:2019rzc} or studying possible modification of gravity \cite{Ezquiaga:2020dao}.
Continuing our studies on gravitational lensing of GWs, and following the results shown here, in a forthcoming paper we will try to assess more precisely the limits of the phase effects, in order to set a more clear limit up to which they are recognizable, as can be seen in Figs.~\ref{fig:NFW2_phases} and \ref{fig:NFWvsNFW2}. We will also extend this study to LIGO types of sources and analyze the potentiality of this method for that specific observational configuration.

%Acknowledgments 
\section*{Acknowledgments}

P.C. is supported by the project ``Uniwersytet 2.0 - Strefa Kariery, ``Miedzynarodowe Studia Doktoranckie Nauk Scislych WMF”, nr POWR.03.05.00-00-Z064/17-00'' co-funded by the European Union through the European Social Fund. DFM thanks the Research Council of Norway for their support, and the resources provided by UNINETT Sigma2 -- the National Infrastructure for High Performance Computing and Data Storage in Norway.

%%%%%%%%%%%%%%%%%%%% REFERENCES %%%%%%%%%%%%%%%%%%

\bibliographystyle{apsrev4-1}
\bibliography{biblio}{}

%%%%%%%%%%%%%%%%% APPENDICES %%%%%%%%%%%%%%%%%%%%%

%\appendix
%\section{Lens models}

\end{document}